\documentclass[journal]{IEEEtran}
\usepackage[cmex10]{amsmath}
\usepackage{amssymb,dsfont,graphicx,tikz}
\usepackage{amsthm}
\usepackage{calc}
\usepackage{amsfonts}
\usepackage[mathscr]{euscript}
\usepackage{float}

\title{Capacity Achieving Codes From Randomness~Condensers} \author{Mahdi Cheraghchi,
  \IEEEmembership{Member, IEEE}

  \thanks{{\footnotesize   Department of Computer Science,
  University of Texas at Austin, TX, USA. Email: $\langle$mahdi@cs.utexas.edu$\rangle$.
      A preliminary
      summary of this work appears (under the title ``Capacity Achieving Codes From Randomness~Conductors'') in
      proceedings of the 2009 IEEE International Symposium on
      Information Theory.
      }}  }

\newtheorem{thm}{Theorem}
\newtheorem{coro}[thm]{Corollary}
\newtheorem{lem}[thm]{Lemma}
\newtheorem{prop}[thm]{Proposition}

\theoremstyle{definition}

\newtheorem{defn}[thm]{Definition}

\newcommand{\F}{\ensuremath{\mathds{F}}}
\newcommand{\N}{\mathds{N}}
\newcommand{\eps}{\epsilon}
\renewcommand{\varepsilon}{\epsilon}

\newcommand{\U}{\mathcal{U}}
\newcommand{\C}{\mathcal{C}}
\newcommand{\Z}{\mathds{Z}}

\newcommand{\supp}{\mathsf{supp}}

\newcommand{\eqdef}{:=}
\newcommand{\cS}{\mathcal{S}}
\newcommand{\cX}{\mathcal{X}}
\newcommand{\cY}{\mathcal{Y}}
\newcommand{\cD}{\mathcal{D}}
\newcommand{\cG}{\mathcal{G}}
\newcommand{\cE}{\mathcal{E}}
\newcommand{\cB}{\mathcal{B}}

\newcommand{\cZ}{\mathcal{Z}}

\newcommand{\zo}{\{0,1\}}

\newcommand{\rk}{{\mathsf{rank}}}
\newcommand{\poly}{{\mathsf{poly}}}
\newcommand{\qpoly}{{\mathsf{quasipoly}}}

\newcommand{\wgt}{\mathsf{wgt}}

\newcommand{\TRASH}[1]{}

\providecommand{\eqref}[1]{(\ref{#1})}

\renewcommand{\qed}{\hfill \IEEEQED}

\usepackage[
     plainpages=false,
     pdftitle={Capacity Achieving Codes From Randomness Condensers},
     pdfsubject={Capacity Achieving Codes From Randomness Condensers},
     pdfstartview=FitH,
     bookmarks=true,
     colorlinks=false,
     bookmarksopen=false,
     bookmarksnumbered=true,
     pdfhighlight=/I,
     hyperfigures=true,
     pdfborder=0 0 0,
     pdfauthor={Mahdi Cheraghchi}]{hyperref}

\providecommand{\cites}[1]{\cite{#1}}

\floatstyle{boxed}
\newfloat{constr}{htb!}{lop}
\floatname{constr}{Construction}
\begin{document}

\setlength{\pdfpageheight}{\paperheight}
\setlength{\pdfpagewidth}{\paperwidth}

\maketitle

\begin{abstract}
  We establish a general framework for construction of small ensembles of
  capacity achieving linear codes for a wide range of (not necessarily
  memoryless) discrete symmetric channels, and in particular, the
  binary erasure and symmetric channels.  The main tool used in our
  constructions is the notion of randomness extractors and lossless
  condensers that are regarded as central tools in theoretical
  computer science.  Same as random codes, the resulting ensembles
  preserve their capacity achieving properties under any change of
  basis.  Using known explicit constructions of condensers,
  we obtain specific ensembles whose size is as small as
  polynomial in the block length.  By applying our construction
  to Justesen's concatenation scheme (Justesen, 1972) we obtain
  explicit capacity achieving codes for BEC (resp., BSC) with almost
  linear time encoding and almost linear time (resp., quadratic time)
  decoding and exponentially small error probability. 

\end{abstract}

\section{Introduction} \label{sec:intro}

\newcommand{\bsc}{\mathsf{BSC}} \newcommand{\bec}{\mathsf{BEC}}
\newcommand{\cF}{\mathcal{F}}

One of the basic goals of coding theory is to come up with efficient
constructions of error-correcting codes that allow reliable
transmission of information over discrete communication
channels. Already in the seminal work of Shannon~\cite{ref:Shannon},
the notion of \emph{channel capacity} was introduced which is a
characteristic of the communication channel that determines the
maximum rate at which reliable transmission of information (i.e., with
vanishing error probability) is possible. However, Shannon's result
did not focus on the \emph{feasibility} of the underlying code and
mainly concerned with the existence of reliable, albeit possibly
complex, coding schemes.  Here feasibility can refer to a combination
of several criteria, including: succinct description of the code and
its efficient computability, the existence of an efficient encoder and
an efficient decoder, the error probability, and the set of message
lengths for which the code is defined and attains its guaranteed
properties.


Besides heuristic attempts, there is a large body of rigorous work in
the literature on coding theory with the aim of designing feasible
capacity approaching codes for various discrete channels, most
notably, the natural and fundamental cases of the binary erasure
channel (BEC) and binary symmetric channel (BSC).  Some notable
examples in ``modern coding'' include Turbo codes and sparse graph
codes (e.g., LDPC codes and Fountain codes, cf.\ \cites{ref:modern,
  ref:blahut, ref:Raptor}).  These classes of codes are either known
or strongly believed to contain capacity achieving ensembles for the
erasure and symmetric channels.

While such codes are very appealing both theoretically and
practically, and are in particular designed with efficient decoding in
mind, in this area there still is a considerable gap between what we
can prove and what is evidenced by practical results, mainly due to
complex combinatorial structure of the code constructions.  Moreover,
almost all known code constructions in this area involve a
considerable amount of randomness, which makes them prone to a
possibility of design failure (e.g., choosing an ``unfortunate''
degree sequence for an LDPC code). While the chance of such
possibilities is typically small, in general there is no known
efficient way to certify whether a particular outcome of the code
construction is satisfactory.  Thus, it is desirable to come up with
constructions of provably capacity achieving code families that are
explicit, i.e., are computationally efficient and do not involve any
randomness.

Explicit construction of capacity achieving codes was considered as
early as the classic work of Forney \cite{ref:forney}, who showed that
concatenated codes can achieve the capacity of various memoryless
channels. In this construction, an outer MDS code is concatenated with
an inner code with small block length that can be found in reasonable
time by brute force search. An important subsequent work by Justesen
\cite{ref:Justesen} (that was originally aimed for explicit
construction of asymptotically good codes) shows that it is possible
to eliminate the brute force search by varying the inner code used for
encoding different symbols of the outer encoding, provided that the
ensemble of inner codes contains a large fraction of capacity
achieving codes.

Recently, Ar{\i}kan~\cite{ref:Arikan09} gave a framework for
deterministic construction of capacity achieving codes for discrete
memoryless channels (DMCs) with binary input that are equipped with
efficient encoders and decoders and attain slightly worse than
exponentially small error probability. These codes are defined for
every block length that is a power of two, which might be considered a
restrictive requirement.  Moreover, the construction is currently
explicit (in the sense of polynomial-time computability of the code
description) only for the special case of BEC and requires exponential
time otherwise.

In this paper, we revisit the concatenation scheme of Justesen and
give new constructions of the underlying ensemble of the inner codes.
The code ensemble used in Justesen's original construction is
attributed to Wozencraft.
Other ensembles that are known to be useful in this scheme include the
ensemble of Goppa codes and shortened cyclic codes
(see~\cite{ref:Roth}, Chapter~12).  The number of codes in these
ensembles is exponential in the block length and they achieve
exponentially small error probability.  These ensembles are also known
to achieve the Gilbert-Varshamov bound \cites{gilbert,varshamov}, and
owe their capacity achieving properties to the property that each
nonzero vector belongs to a small number of the codes in the ensemble.

In this work, we will use randomness extractors and lossless
condensers that are fundamental objects in theoretical computer
science and in particular, theory of pseudorandomness, to construct
much smaller ensembles with similar, random-like, properties.  The
quality of the underlying extractor or condenser determines the
quality of the resulting code ensemble. In particular, the size of the
code ensemble, the decoding error and proximity to the channel
capacity are determined by the \emph{seed length}, the \emph{error},
and the \emph{output length} of the extractor or condenser being used.




As a concrete example, we will instantiate our construction with
appropriate choices of the underlying condenser (or extractor) and
obtain, for every (sufficiently large) block length $n$, a capacity
achieving ensemble of size $2^n$ that attains exponentially small
error probability for both erasure and symmetric channels (as well as
the broader range of channels described above), and an ensemble of
quasipolynomial\footnote{A quantity $f(n)$ is said to be
  quasipolynomial\index{quasipolynomial} in $n$ (denoted by $f(n) =
  \qpoly(n)$) if $f(n) = 2^{(\log n)^{O(1)}}$.} size $2^{O(\log^3 n)}$
that attains the capacity of BEC.  Using nearly optimal extractors and
condensers that require logarithmic seed lengths, it is possible to
obtain polynomially small capacity achieving ensembles for any block
length and this is what we achieve from a lossless condenser due to
Guruswami, Umans, and Vadhan~\cite{ref:GUV09}.

Finally, we apply our constructions to
Justesen's concatenation scheme to obtain an explicit construction of
capacity-achieving codes for both BEC and BSC that attain
exponentially small error, as in the original construction of Forney.
Moreover, the running time of the encoder is almost linear in the
block length, and decoding takes almost linear time for BEC and almost
quadratic time for BSC. Using our subexponential-sized ensembles as
the inner code, we are able to construct explicit codes for BEC and
BSC that are defined and capacity achieving for every choice of the
message length.

The rest of the paper is organized as follows. In
Section~\ref{sec:prelim}, we review some basic definitions and facts
on discrete communication channels\footnote{ A detailed treatment of
  this topic can be found in the book by Cover and Thomas
  \cite{ref:cover}.}  and introduce the notion of randomness
extractors and condensers that are our main technical tools.  In
Section~\ref{sec:bec} we construct our code ensembles for the binary
erasure channel. This is followed by the code ensembles for the binary
symmetric and additive noise channels in Section~\ref{sec:bsc}.  In
Section~\ref{sec:explicit} we incorporate our code ensembles into a
code concatenation scheme and construct explicit capacity achieving
codes.  Finally, Section~\ref{sec:dualityAffine} proves a duality
theorem for linear affine extractors and condensers that can be seen
as a byproduct of the techniques used in the paper, with potential
applications of independent interest in the theory of randomness
extractors.

\section{Preliminaries} \label{sec:prelim}

\subsection{Discrete Communication Channels} \label{sec:dmc}

\newcommand{\SC}{\mathsf{SC}} \newcommand{\CAP}{\mathsf{Cap}}
\newcommand{\Ch}{\mathscr{C}}

A \emph{discrete communication channel}\index{channel} is a randomized
process that takes a potentially infinite stream of symbols $X_0, X_1,
\ldots$ from an \emph{input alphabet} $\Sigma$ and outputs an infinite
stream $Y_0, Y_1, \ldots$ from an \emph{output alphabet} $\Gamma$.
The indices intuitively represent the \emph{time}, and each output
symbol is only determined from what channel has observed in the
past. More precisely, given $X_0, \ldots, X_t$, the output symbol
$Y_t$ must be independent of $X_{t+1}, X_{t+2}, \ldots$. In this work,
we will concentrate on finite input and finite output channels, that
is, when the alphabets $\Sigma$ and $\Gamma$ are finite. In this case,
at every time instance $t \in \N \cup \{0\}$, the conditional distribution $p(Y_t|X_t)$
of each output symbol $Y_t$ given the input symbol $X_t$ (and the past
outcomes) can be written as a stochastic $|\Sigma| \times |\Gamma|$
\emph{transition matrix}, where each row is a probability
distribution.

Of particular interest is a \emph{memoryless}
channel\index{channel!memoryless}, which is intuitively ``oblivious''
of the past. In this case, the transition matrix is independent of the
time instance. That is, we have $p(Y_t | X_t)=p(Y_0 | X_0)$ for every
$t \in \N$. When the rows of the transition matrix are permutations of one
another and so is the case for the columns, the channel is called
\emph{symmetric}\index{channel!symmetric}. For example, the channel
defined by
\[
p(Y|X) = \begin{pmatrix} 0.4 & 0.1 & 0.5 \\ 0.5 & 0.4 & 0.1 \\ 0.1 &
  0.5 & 0.4
\end{pmatrix}
\]
is symmetric. Intuitively, a symmetric channel does not ``read'' the
input sequence.  An important class of symmetric channels is defined
by \emph{additive noise}.  In an additive noise channel, the input and
output alphabets are the same finite field $\F_q$ and each output
symbol $Y_t$ is obtained from $X_t$ using
\[
Y_t = X_t + Z_t,
\]
where the addition is over $\F_q$ and the channel noise $Z_t \in \F_q$
is chosen independently of the input sequence\footnote{ In fact, since
  we are only using the additive structure of $\F_q$, it can be
  replaced by any additive group, and in particular, the ring $\Z/q\Z$
  for an arbitrary integer $q>1$. This way, $q$ does not need to be
  restricted to be a prime power.}.  Typically $Z_t$ is also independent
of time $t$, in which case we get a memoryless additive noise channel.
For a noise distribution $\cZ$, we denote the memoryless additive
noise channel over the input (as well as output) alphabet $\Sigma$ by
$\SC(\Sigma, \cZ)$.

Note that the notion of additive noise channels can be extended to the
case where the input and alphabet sets are vector spaces $\F_q^n$, and
the noise distribution is a probability distribution over $\F_q^n$. By
considering an isomorphism between $\F_q^n$ and the field extension
$\F_{q^n}$, such a channel is essentially an additive noise channel
$\SC(\F_{q^n}, \cZ)$, where $\cZ$ is a noise distribution over
$\F_{q^n}$. On the other hand, the channel $\SC(\F_{q^n}, \cZ)$ can be
regarded as a ``block-wise memoryless'' channel over the alphabet
$\F_q$. Namely, in a natural way, each channel use over the alphabet
$\F_{q^n}$ can be regarded as $n$ subsequent uses of a channel over
the alphabet $\F_q$. When regarding the channel over $\F_q$, it does
not necessarily remain memoryless since the additive noise
distribution $\cZ$ can be an arbitrary distribution over $\F_{q^n}$
and is not necessarily expressible as a product distribution over
$\F_q$. However, the noise distributions of blocks of $n$ subsequent
channel uses are independent from one another and form a product
distribution (since the original channel $\SC(\F_{q^n}, \cZ)$ is
memoryless over $\F_{q^n}$).  Often by choosing larger and larger
values of $n$ and letting $n$ grow to infinity, it is possible to
obtain good approximations of a non-memoryless additive noise channel
using memoryless additive noise channels over large alphabets.

An important additive noise channel is the \emph{$q$-ary symmetric
  channel}\index{channel!$q$-ary symmetric}, which is defined by a
(typically small) noise parameter $p \in [0, 1-1/q)$. For this channel,
the noise distribution $\cZ$ has a probability mass $1-p$ on zero, and
$p/(q-1)$ on every nonzero alphabet letter.  A fundamental special
case is the \emph{binary symmetric channel}\index{channel!binary
  symmetric (BSC)} (BSC), which corresponds to the case $q=2$ and is
denoted by $\bsc(p)$.

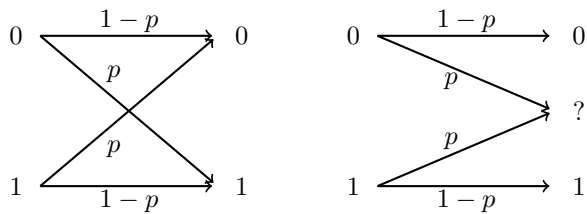
\begin{figure}[t]
  \begin{center}
    \mbox{
      \begin{tikzpicture}[rounded corners, thick, scale=1]
        \path (0,2) node[circle] (left_0) {$0$} (0,0) node[circle]
        (left_1) {$1$} (3,0) node[circle] (right_1) {$1$} (3,2)
        node[circle] (right_0) {$0$};
  
        \draw [->,shorten >= 2pt] (left_0.east) -- (right_0.west);
        \draw [->,shorten >= 2pt] (left_0.east) -- (right_1.west);
        \draw [->,shorten >= 2pt] (left_1.east) -- (right_1.west);
        \draw [->,shorten >= 2pt] (left_1.east) -- (right_0.west);

        \draw (1.5,2.2) node {$1-p$}; \draw (1.5,-0.2) node {$1-p$};
        \draw (1.3,1.5) node {$p$}; \draw (1.3,0.5) node {$p$};
      \end{tikzpicture}} \hspace{5mm} \mbox{
      \begin{tikzpicture}[rounded corners, thick, scale=1]
        \path (0,2) node[circle] (left_0) {$0$} (0,0) node[circle]
        (left_1) {$1$} (3,0) node[circle] (right_1) {$1$} (3,2)
        node[circle] (right_0) {$0$} (3,1) node[circle] (right_E)
        {$?$};
  
        \draw [->,shorten >= 2pt] (left_0.east) -- (right_0.west);
        \draw [->,shorten >= 2pt] (left_0.east) -- (right_E.west);
        \draw [->,shorten >= 2pt] (left_1.east) -- (right_1.west);
        \draw [->,shorten >= 2pt] (left_1.east) -- (right_E.west);

        \draw (1.5,2.2) node {$1-p$}; \draw (1.5,-0.2) node {$1-p$};
        \draw (1.3,1.4) node {$p$}; \draw (1.3,0.6) node {$p$};
      \end{tikzpicture}}
  \end{center}
  \caption[The binary symmetric and binary erasure channels]{The
    binary symmetric channel (left) and binary erasure channel
    (right). On each graph, the left part corresponds to the input
    alphabet and the right part to the output alphabet. Conditional
    probability of each output symbol given an input symbol is shown
    by the labels on the corresponding arrows.}
  \label{fig:dmc}
\end{figure}

Another fundamentally important channel is the \emph{binary erasure
  channel}\index{channel!binary erasure (BEC)}.  The input alphabet
for this channel is $\{0,1\}$ and the output alphabet is the set
$\{0,1,?\}$.  The transition is characterized by an \emph{erasure
  probability} $p \in [0,1)$. A transmitted symbol is output intact by
the channel with probability $1-p$. However, with probability $p$, a
special \emph{erasure symbol} ``$?$'' is delivered by the channel. The
behavior of the binary symmetric channel $\bsc(p)$ and binary erasure
channel $\bec(p)$ is schematically described by Fig.~\ref{fig:dmc}.

A \emph{channel encoder} $\mathcal{E}$ for a channel $\mathscr{C}$
with input alphabet $\Sigma$ and output alphabet $\Gamma$ is a mapping
$\C\colon \zo^k \to \Sigma^n$. A \emph{channel decoder}, on the other
hand, is a mapping $\mathcal{D}\colon \Gamma^n \to \zo^k$. A channel
encoder and a channel decoder collectively describe a \emph{channel
  code}. Note that the image of the encoder mapping defines a block
code\footnote{We refer the reader to the books by
MacWilliams and Sloane \cite{ref:MS}, van~Lint \cite{ref:vanl}, and
Roth \cite{ref:Roth} for the basic notions in coding theory.} 
of length $n$ over the alphabet $\Sigma$. The parameter $n$
defines the \emph{block length} of the code.  For a sequence $Y \in
\Sigma^n$, denote by the random variable $\mathscr{C}(Y)$ the sequence
$\hat{Y} \in \Gamma^n$ that is output by the channel, given the input
$Y$.

\begin{figure*}
  \begin{center} \vspace{5mm}
    \includegraphics[width=\textwidth]{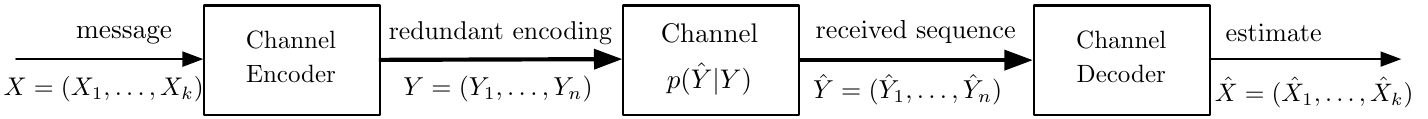}
  \end{center}
  \caption[The schematic diagram of a point-to-point communication
  system]{The schematic diagram of a point-to-point communication
    system. The stochastic behavior of the channel is captured by the
    conditional probability distribution $p(\hat{Y}|Y)$.}
  \label{fig:channel}
\end{figure*}

Intuitively, a channel encoder adds sufficient redundancy to a given
``message'' $X \in \zo^k$ (that is without loss of generality modeled
as a binary string of length $k$), resulting in an encoded sequence $Y
\in \Sigma^n$ that can be fed into the channel. The channel
manipulates the encoded sequence and delivers a sequence $\hat{Y} \in
\Gamma^n$ to a recipient whose aim is to recover $X$.  The recovery
process is done by applying the channel decoder on the received
sequence $\hat{Y}$.  The transmission is successful when
$\mathcal{D}(\hat{Y})=X$. Since the channel behavior is not
deterministic, there might be a nonzero probability, known as the
\emph{error probability}, that the transmission is unsuccessful. More
precisely, the error probability of a channel code is defined as
\[
p_e := \sup_{X \in \zo^k} \Pr[\mathcal{D}(\mathscr{C}(\mathcal{E}(X)))
\neq X],
\]
where the probability is taken over the randomness of $\mathscr{C}$.
A schematic diagram of a simple communication system consisting of an
encoder, point-to-point channel, and decoder is shown in
Fig.~\ref{fig:channel}.

For linear codes over additive noise channels, it is often convenient
to work with \emph{syndrome decoders}. Consider a linear code with
generator and parity check matrices $G$ and $H$, respectively. The
encoding of a message $x$ (considered as a row vector) can thus be
written as $xG$.  Suppose that the encoded sequence is transmitted
over an additive noise channel, which produces a noisy sequence $y :=
xG+z$, for a randomly chosen $z$ according to the channel
distribution.  The receiver receives the sequence $y$ and, without
loss of generality, the decoder's task is to obtain an estimate of the
noise realization $z$ from $y$. Now, observe that
\[
Hy^\top = HG^\top x^\top + H z^\top = H z^\top,
\]
where the last equality is due to the orthogonality of the generator
and parity check matrices. Therefore, $H z^\top$ is available to the
decoder and thus, in order to decode the received sequence, it
suffices to obtain an estimate of the noise sequence $z$ from the
\emph{syndrome} $H z^\top$. A syndrome decoder is a function that,
given the syndrome, outputs an estimate of the noise sequence (note
that this is independent of the codeword being sent). The error
probability of a syndrome decoder can be simply defined as the
probability (over the noise randomness) that it obtains an incorrect
estimate of the noise sequence. Obviously, the error probability of a
syndrome decoder upper bounds the error probability of the channel
code.

The \emph{rate} of a channel code (in bits per channel use) is defined
as the quantity $k/n$.  We call a rate $r \geq 0$ \emph{feasible} if
for every $\eps > 0$, there is a channel code with rate $r$ and error
probability at most $\eps$. The rate of a channel code describes its
efficiency; the larger the rate, the more information can be
transmitted through the channel in a given ``time frame''.  A
fundamental question is, given a channel $\mathscr{C}$, to find the
largest possible rate at which reliable transmission is possible. In
his fundamental work, Shannon \cite{ref:Shannon} introduced the notion
of \emph{channel capacity} that answers this question.  Shannon
capacity can be defined using purely infor\-mation-theoretic
terminology. However, for the purposes of this paper, it is more
convenient to use the following, more ``computational'', definition
which turns out to be equivalent to the original notion of Shannon
capacity:
\[
\CAP(\mathscr{C}) := \sup \{ r \mid \text{$r$ is a feasible rate for
  the channel $\mathscr{C}$} \}.
\]

Capacity of memoryless symmetric channels has a particularly nice
form. Let $\cZ$ denote the probability distribution defined by any of
the rows of the transition matrix of a memoryless symmetric channel
$\mathscr{C}$ with output alphabet $\Gamma$. Then, capacity of
$\mathscr{C}$ is given by
\[
\CAP(\mathscr{C}) = \log_2 |\Gamma| - H(\cZ),
\]
where $H(\cdot)$ denotes the Shannon entropy
\cite[Section~7.2]{ref:cover}. In particular, capacity of the binary
symmetric channel $\bsc(p)$ (in bits per channel use) is equal to
\[
1-h(p) = 1+p \log_2 p +(1-p) \log_2 (1-p).
\]
Moreover, capacity of the binary erasure channel $\bec(p)$ is known to
be $1-p$ \cite[Section~7.1]{ref:cover}.

A \emph{family}\index{code!family} of channel codes of rate $r$ is an
infinite set of channel codes, such that for every (typically small)
\emph{rate loss} $\delta \in (0, r)$ and block length $n$, the family
contains a code $\C(n, \delta)$ of length at least $n$ and rate at
least $r-\delta$. The family is called \emph{explicit} if there is a
deterministic algorithm that, given $n$ and $\delta$ as parameters,
computes the encoder function of the code $\C(n, \delta)$ in
polynomial time in $n$.  For linear channel codes, this is equivalent
to computing a generator or parity check matrix of the code in
polynomial time.  If, additionally, the algorithm receives an
auxiliary index $i \in [s]$, for a \emph{size parameter} $s$ depending
on $n$ and $\delta$, we instead get an
\emph{ensemble}\index{code!ensemble} of size $s$ of codes. An ensemble
can be interpreted as a set of codes of length $n$ and rate at least
$r-\delta$ each, that contains a code for each possibility of the
index $i$.

We call a family of codes \emph{capacity
  achieving}\index{code!capacity achieving} for a channel
$\mathscr{C}$ if the family is of rate $\CAP(\mathscr{C})$ and
moreover, the code $\C(n, \delta)$ as described above can be chosen to
have an arbitrarily small error probability for the channel
$\mathscr{C}$.  If the error probability decays exponentially with the
block length $n$; i.e., $p_e = O(2^{-\gamma n})$, for a constant
$\gamma > 0$ (possibly depending on the rate loss), then the family is
said to achieve an \emph{error exponent}\index{code!error exponent}
$\gamma$.  We call the family \emph{capacity achieving for all
  lengths} if it is capacity achieving and moreover, there is an
integer constant $n_0$ (depending only on the rate loss $\delta$) such
that for every $n \geq n_0$, the code $\C(n, \delta)$ can be chosen to
have length exactly $n$.

\subsection{Extractors and Condensers} \label{sec:extr}

Extractors and
condensers 
are basic notions in theoretical computer science (and in particular
derandomization theory) that constitute the main technical tools that
we use for our code constructions. In this section, we review the
essential notions and tools related to these objects and probability
distributions on finite sample spaces. A detailed treatment of
derandomization theory can be found, among other sources, in
\cite{ref:AB09}.

The \emph{min-entropy} of a probability distribution $\cX$ over a
finite sample space with support\footnote{Support of a distribution
  $\cX$ is the set of points of the sample space to which $\cX$
  assigns nonzero probability.} $S$ (in symbols, $S := \supp(\cX)$) is
given by \[
\min_{x \in S}\{-\log \Pr_\cX(x)\}, \] where $\Pr_\cX(x)$ is the
probability that $\cX$ assigns to $x$.  All unsubscripted logarithms
in this work are to the base $2$ (and thus, the entropy is measured in
bits).  For a distribution $\cX$ over $\F_2^n$, the \emph{entropy
  rate} of $\cX$ is given by $H_\infty(\cX)/n$.

The \emph{statistical distance} between two distributions $\cX$ and
$\cY$ defined on the same finite space $\Omega$ is given by \[ \frac{1}{2}
\sum_{s \in \Omega} |\Pr_\cX(s) - \Pr_\cY(s)|, \] which is half the
$\ell_1$ distance of the two distributions when regarded as vectors of
probabilities over $\Omega$. Two distributions $\cX$ and $\cY$ are said to
be $\eps$-close if their statistical distance is at most $\eps$.  We
will use the shorthand $\U_n$ for the uniform distribution on
$\F_2^n$, where $\F_2$ denotes the finite field over $\zo$,
$\U_S$ for the uniform distribution on a finite set $S$, and
$X \sim \cX$ to denote a random variable $X$ drawn from a distribution
$\cX$.


We will occasionally consider \emph{convex combinations} of
distributions, defined as follows:

\begin{defn}
  Let $\cX_1, \cX_2, \ldots, \cX_t$ be probability distributions over
  a finite space $\Omega$ and $\alpha_1, \alpha_2, \ldots, \alpha_t$
  be nonnegative real values that sum up to $1$. Then the convex
  combination \[ \alpha_1 \cX_1 + \alpha_2 \cX_2 + \cdots + \alpha_t
  \cX_t \] is a distribution $\cX$ over $\Omega$ given by the
  probability measure
  \[ \Pr_\cX(x) \eqdef \sum_{i=1}^t \alpha_i \Pr_{\cX_i}(x), \] for
  every $x \in \Omega$.
\end{defn}

A \emph{flat} distribution is a distribution that is uniform on its
support.  For flat distributions, min-entropy coincides with the
Shannon-entropy.  The set of probability distributions with
min-entropy at least $m$ forms a convex set with vertices
corresponding to flat distribution with min-entropy $m$ (when $2^m$ is
an integer). Therefore, any distribution with min-entropy at least $m$
can be written as a convex combination
of flat distributions with min-entropy $m$.  We will use the following
proposition regarding flat probability distributions (a proof is
presented in Appendix~\ref{app:flatmap}):

\begin{prop} \label{prop:flatmap} Let $\cX$ be a flat distribution
  with min-entropy $\log M$ over a finite sample space $\Omega$ and
  $f\colon \Omega \to \Gamma$ be a mapping to a finite set $\Gamma$.
  \begin{enumerate}
  \item If $f(\cX)$ is $\eps$-close to having min-entropy $\log M$,
    then there is a set $T \subseteq \Gamma$ of size at least
    $(1-2\eps)M$ such that
    \begin{multline*}
      (\forall y \in T\text{ and }\forall x,x'\in \supp(\cX))\colon \\
      f(x)=y \land f(x') = y \Rightarrow x = x'.
    \end{multline*}
  \item Suppose $|\Gamma| \geq M$. If $f(\cX)$ has a support of size
    at least $(1-\eps)M$, then it is $\eps$-close to having
    min-entropy $\log M$.
  \end{enumerate} \qed
\end{prop}

The general notion of randomness condensers that we use in this work
is the following:

\begin{defn} \index{condenser} \label{def:condenser} A function
  $f\colon \F_q^n \times \zo^d \to \F_q^r$ is an \emph{$m \to_\eps m'$
    condenser\footnote{In the extractor theory literature, the
      definition presented here corresponds to \emph{strong}
      condensers, as opposed to regular (weaker) condensers.  However
      since we will only deal with strong condensers in this work, we
      omit the word ``strong'' throughout.}} if for every distribution
  $\cX$ on $\F_q^n$ with min-entropy at least $m$, random variable $X
  \sim \cX$ and a \emph{seed} $Y \sim \U_d$, the distribution of $(Y,
  f(X, Y))$ is $\eps$-close to a distribution $(\U_d, \cZ)$ with
  min-entropy at least $d+m'$. The parameters $m$, $m'$, $\eps$, $m -
  m'$, and $r \log q - m'$ are called the \emph{input entropy},
  \emph{output entropy}, \emph{error}, the \emph{entropy loss} and the
  \emph{overhead} of the condenser, respectively.  A condenser is
  \emph{explicit} if it is polynomial-time computable, and
  \emph{linear} if it is a linear function in the first argument.
  That is, when for every fixed seed $z \in \zo^d$, every $x, x' \in
  \F_q^n$ and scalars $\alpha, \beta \in \F_q$ we have
  \[
  f(\alpha x+ \beta x', z) = \alpha f(x,z) + \beta f(x',z),
  \]
  where the addition is coordinate-wise and over the field $\F_q$.
\end{defn}

In this paper, for the most part, we deal with the binary case, where
$q=2$.  Two extremal cases of Definition~\ref{def:condenser} are of
particular interest.  Namely, when the overhead $r \log q - m'$ is
zero, the output distribution is close to uniform and the condenser is
called an \emph{extractor}. On the other hand, when the entropy loss
$m - m'$ is zero, the condenser does not lose any of the source
entropy and is therefore called a \emph{lossless condenser}.  We will
use the shorthand $(m, \eps)$-extractor for an $m \to_\eps m'$
condenser with output length $m'$ and $(m, \eps$)-lossless condenser
for an $m \to_\eps m$ condenser. For technical reasons, in this work
we additionally require $(m, \eps)$-lossless condensers to remain
$(m', \eps)$-lossless condensers for every $m' \leq m$. All the
explicit constructions that we mention and use satisfy this property.

When we have a particular random source in mind, a seedless variation
of Definition~\ref{def:condenser} as follows often becomes useful.

\begin{defn}
  A function $f\colon \F_q^n \to \F_q^m$ is an $m \to_\eps m'$
  condenser for a particular source $\cX$ having min-entropy at least
  $m$ if the distribution $f(\cX)$ is $\eps$-close to a distribution
  that has min-entropy at least $m'$.
\end{defn}

The extremal notions of extractors and lossless condensers naturally
extend to the seedless definition as well. A standard averaging
argument can show that a seeded condenser is in fact an equally good
condenser for any source for almost every fixing of the seed. More
precisely, we have the following:

\begin{prop} \label{prop:strongExt} Let $f\colon \F_q^n \times \zo^d
  \to \F_q^r$ be an $m \to_\eps m'$ condenser.  Consider an arbitrary
  parameter $\delta > 0$ and a source $\cX$ with min-entropy $m$ or
  more.  Then, for all but at most a $\delta$ fraction of the choices
  of $z \in \zo^d$, the function $f(\cdot, z)$ is an $m
  \to_{\eps/\delta} m'$ condenser for $\cX$. \qed
\end{prop}

A seedless function may simultaneously be a condenser for a family of
sources. Two important families that we consider in this work are
\emph{bit-fixing} and \emph{affine} sources. A bit-fixing source of
length $n$ with min-entropy $m$ is a distribution of random variables
$(X_1, \ldots X_n) \in \F_2^n$ such that for some (unknown) set $S
\subseteq [n]$ of size $m$, the variables $(X_i\colon i \in S)$ are
independent and uniformly distributed, and the remaining variables are
fixed to arbitrary values. More generally, an \emph{affine} source
over $\F_q^n$ of min-entropy $m \log q$ is a flat distribution
supported on an affine translation of some (unknown) $m$-dimensional
vector space in $\F_q^n$. An $m \to_\eps m'$ bit-fixing (resp., affine)
condenser is a function that is an $m \to_\eps m'$ condenser for any
bit-fixing (resp., affine) source with min-entropy at least $m$.  Same
as before, important special cases include bit-fixing, or affine,
extractors and lossless condensers.

The code ensembles that we are going to use in this work are based on
\emph{linear} extractors and lossless condensers.  In order to get
capacity-achieving code ensembles, we need condensers whose output
length $r$ is very close to the input entropy $m$; namely, $|r-m| \leq
\alpha m$ for an arbitrarily small constant $\alpha$.  Moreover, in
order to get explicit code ensembles, we need to instantiate our
constructions with explicit condensers. Explicit constructions of
condensers that are best suitable for our applications include
condensers from the \emph{Left-over Hash Lemma}, Trevisan's extractor,
and the lossless condenser of Guruswami, Umans, and Vadhan. These
constructions are briefly reviewed below.

\subsubsection{The Leftover Hash Lemma}


One of the foremost explicit constructions of linear extractors is
given by the \emph{Leftover Hash Lemma} first stated by Impagliazzo,
Levin, and Luby \cite{ref:ILL89}. This extractor achieves an optimal
output length $r = m - 2\log(1/\eps)$ albeit with a substantially
large seed length $d = n$.  In its general form, the lemma states that
any \emph{universal family of hash functions} can be transformed into
an explicit extractor. The universality property required by the hash
functions is captured by the following definition.

\newcommand{\cH}{\mathcal{H}}
\newcommand{\hLin}{\mathcal{H}_{\mathsf{lin}}}

\begin{defn} \label{defn:pwindep} \index{universal hash family}
  A family of functions $\cH = \{h_1, \ldots, h_D\}$ where $h_i\colon
  \F_2^n \to \F_2^r$ for $i=1, \ldots, D$ is called \emph{universal}
  if, for every fixed choice of $x, x' \in \F_2^n$ such that $x \neq
  x'$ and a uniformly random $i \in [D] := \{1, \ldots,
  D\}$\index{notation!$[n] := \{1,\ldots,n\}$} we have
  \[
  \Pr_i [h_i(x) = h_i(x')] \leq 2^{-r}.
  \]
\end{defn}

One of the basic examples of universal hash families is what we call
\emph{the linear family}, defined as follows. Consider an arbitrary
isomorphism $\varphi\colon \F_2^n \to \F_{2^n}$ between the vector
space $\F_2^n$ and the extension field $\F_{2^n}$, and let $0 < r \leq
n$ be an arbitrary integer.  The linear family $\hLin$ is the set $\{
h_\alpha\colon \alpha \in \F_{2^n} \}$ of size $2^n$ that contains a
function for each element of the extension field $\F_{2^n}$.  For each
$\alpha$, the mapping $h_\alpha$ is given by
\[
h_\alpha(x) := (y_1, \ldots, y_r), \text{ where $(y_1, \ldots, y_n) :=
  \varphi^{-1}(\alpha \cdot \varphi(x))$}.
\]
Observe that each function $h_\alpha$ can be expressed as a linear
mapping from $\F_2^n$ to $\F_2^r$.  It is well known and
straightforward to see that this family is indeed universal (cf.\
\cite{ref:MR}).



The following is a straightforward generalization of the Leftover Hash
Lemma which shows that universal hash families can be used to
construct not only extractors, but also lossless condensers.  For
completeness, we present a proof of this extension of this extension
(which is not much different from the original proof) in
Appendix~\ref{app:LOHL}.

\begin{lem} (Leftover Hash Lemma) \index{Leftover Hash
    Lemma} \label{lem:leftover} Let $\cH = \{ h_i\colon \F_2^n \to
  \F_2^r \}_{i \in \F_2^d}$ be a universal family of hash functions
  with $2^d$ elements indexed by binary vectors of length $d$, and
  define the function $ f\colon \F_2^n \times \F_2^d \to \F_2^r $ as
  $f(x, z) := h_z(x)$. Then
  \begin{enumerate}
  \item For every $m, \eps$ such that $r \leq m - 2 \log(1/\eps)$, the
    function $f$ is an $(m,\eps)$-extractor, and
 
  \item For every $m, \eps$ such that $r \geq m + 2 \log(1/\eps)$, the
    function $f$ is a lossless $(m,\eps)$-condenser.
  \end{enumerate}
  In particular, by choosing $\cH = \hLin$, it is possible to get
  explicit extractors and lossless condensers with seed length $d=n$.
\end{lem}

\subsubsection{Trevisan's Extractor}

Another basic example of an explicit extractor is Trevisan's
construction \cite{ref:Tre}. The extractor is based on ``black-box
pseudorandom generators'' and works by encoding the input using a
linear error-correcting code and then carefully puncturing the
outcome. Clearly an extractor constructed this way is linear. An
improved analysis of the extractor due to Raz et al.~results in the
following theorem:

\begin{thm} \cite{ref:RRV} \label{thm:RRV} For all positive integers
  $n, m$ and real parameter $\eps > 0$, there is an explicit linear
  $(m, \eps)$-extractor $g\colon \F_2^n \times \F_2^{d} \to \F_2^r$
  with $d = O(\log^3 (n/\eps))$ and $r = m - O(d)$. \qed
\end{thm}

\subsubsection{Lossless Condenser of Guruswami-Umans-Vadhan}

To this date, one of the best constructions of lossless condensers for
the range of parameters of our interest is due to Guruswami et
al.~\cite{ref:GUV09}.  The condenser is based on error-correcting
codes and can be best described as a mapping $\F_q^{\bar{n}} \times
\F_q \to \F_q^{\bar{r}}$ over a finite field $\F_q$ (which for us
would be an extension of $\F_2$). The mapping is depicted in
Construction~\ref{fig:GUV}. It can be proved \cite{ref:GUV09} that this
construction indeed gives a lossless condenser for an appropriate
setting of the parameters. Even though in general the mapping is not
linear, we observe that for any subfield $\F_{\bar{q}}$ of $\F_q$, the
parameter $h$ can be chosen to be an integer power of ${\bar{q}}$. It
is straightforward to verify that this restriction does not affect the
quality of the condenser\footnote{This is true as long as ${\bar{q}}$
  remains a fixed constant.}, and furthermore, makes the mapping
$F_{\bar{q}}$-linear for every fixed seed.  Thus, using a standard
embedding of $\F_q$ as an $\F_{\bar{q}}$-linear vector space, the
condenser can be regarded as a linear function over $\F_{\bar{q}}$.

The following result (Theorem~\ref{thm:GUV}), proved in
\cite{ref:GUV09}, summarizes the parameters of the condenser. Combined
with the above observation, we can also ensure that the condenser is
linear. In particular, for the range of parameters that is of interest
in this work, the result gives an $\F_2$-linear $(m,\eps$)-lossless
condenser with logarithmic seed length $d=O(\log n)$ and output length
$r \leq (1+\alpha)m$, where $\alpha$ is an arbitrarily small positive
constant.

\newcommand{\barN}{{\bar{n}}} \newcommand{\barR}{{\bar{r}}}
\begin{thm} \label{thm:GUV} Let $q$ be a fixed prime power and $\alpha
  > 0$ be an arbitrary constant. Then, for parameters $\barN \in \N$,
  $m \leq \barN \log q$, and $\eps > 0$, there is an explicit $(m,
  \eps)$-lossless condenser $f\colon \F_{q}^\barN \times \zo^d \to
  \F_{q}^\barR$ with seed length $d \leq (1+1/\alpha) (\log (\barN
  m/\eps) + O(1))$ and output length satisfying $\barR \log q \leq
  d+(1+\alpha)m$. Moreover, $f$ is a linear function (over $\F_{q}$)
  for every fixed choice of the seed. \qed
\end{thm}

\newcommand{\GUV}{\mathsf{GUV}}
\begin{constr}
  \begin{itemize}
  \item \textit{Given:} A random sample $X \sim \cX$, where $\cX$ is a
    distribution on $\F_q^{\bar{n}}$ with min-entropy at least $m$,
    and a uniformly distributed random seed $Z \sim \U_{\F_q}$ over
    $\F_q$.

  \item \textit{Output:} A vector $\GUV(X,Z)$ of length $\bar{r}$ over
    $\F_q$.

  \item \textit{Construction:} Take any irreducible univariate
    polynomial $G$ of degree $\bar{n}$ over $\F_q$, and interpret the
    input $X$ as the coefficient vector of a random univariate
    polynomial $F$ of degree $\bar{n}-1$ over $\F_q$. Then, for an
    integer parameter $h$, the output is given by
    \[
    \GUV(X,Z) := (F(Z), F_1(Z), \ldots, F_{\bar{r}-1}(Z)),
    \]
    where we have used the shorthand $F_i := F^{h^i} \mod G$.
    \caption{Guruswami-Umans-Vadhan's Condenser $\GUV\colon
      \F_q^{\bar{n}} \times \F_q \to \F_q^{\bar{r}}$.}
    \label{fig:GUV}
  \end{itemize}
\end{constr}


\section{Codes for the Binary Erasure Channel} \label{sec:bec}

Any code with minimum distance $d$ can tolerate up to $d-1$ erasures
in the \emph{worst case}. 
Thus, one way to ensure reliable communication over $\bec(p)$ is to use
binary codes with relative minimum distance of about $p$. However,
known negative bounds on the rate-distance trade-off (e.g., the sphere
packing and MRRW bounds) do not allow the rate of such codes to
approach the capacity $1-p$.  However, by imposing the weaker
requirement that \emph{most} of the erasure patterns should be
recoverable, it is possible to attain the capacity with a positive,
but arbitrarily small, error probability (as guaranteed by the
definition of capacity).

In this section, we consider a different relaxation that preserves the
worst-case guarantee on the erasure patterns; namely we consider
\emph{ensembles} of linear codes with the property that \emph{any}
pattern of up to $p$ erasures must be tolerable by all but a
negligible fraction of the codes in the ensemble. This in particular
allows us to construct ensembles in which all but a negligible
fraction of the codes are capacity achieving for BEC.  We remark that since we
are only considering linear codes, recoverability from a particular
erasure pattern $S \subseteq [n]$ (where $n$ is the block length)
is a property of the code and independent of the encoded sequence.

Now we introduce two constructions, which employ linear
extractors and lossless condensers as their main
ingredients. Throughout this section we denote by $f\colon \F_2^n
\times \F_2^d \to \F_2^r$ a linear, lossless condenser for
min-entropy $m$ and error $\eps$ and by $g\colon \F_2^n \times
\F_2^{d'} \to \F_2^k$ a linear extractor for min-entropy $n-m$
and error $\eps'$. We assume that the errors $\eps$ and $\eps'$ are
substantially small.  Using this notation, we define the ensembles
$\cF$ and $\cG$ as in Construction~\ref{constr:ensembles}.

\begin{constr} 
    \begin{itemize}
    \item \emph{Ensemble $\cF$:} Define a code $\C_u$ for each seed $u \in
      \F_2^d$ as follows: Let $H_u$ denote the $r \times n$ matrix
      that defines the linear function $f(\cdot, u)$, i.e., for each
      $x \in \F_2^n$, $H_u \cdot x = f(x,u)$. Then $H_u$ is a parity
      check matrix for $\C_u$.

    \item \emph{Ensemble $\cG$:} Define a code $\C'_u$ for each seed $u \in
      \F_2^{d'}$ as follows: Let $G_u$ denote the $k \times n$ matrix
      that defines the linear function $g(\cdot, u)$.  Then $G_u$ is a
      generator matrix for $\C'_u$.
    \end{itemize}
  \caption{Ensembles $\cF$ and $\cG$ of error-correcting codes.}
  \label{constr:ensembles}
\end{constr}

Obviously, the rate of each code in $\cF$ is at least
$1-r/n$. Moreover, we can assume without
loss of generality that the rank of each $G_u$ is exactly\footnote{
  This causes no loss of generality since, if the rank of some $G_u$
  is not maximal, one of the $k$ symbols output by the linear function
  $g(\cdot, u)$ would linearly depend on the others and thus, the
  function would fail to be an extractor for any source (so one can
  arbitrarily modify $g(\cdot, u)$ to have rank $k$ without negatively
  affecting the parameters of the extractor $g$).  } $k$. Thus, each
code in $\cG$ has rate $k/n$. Lemma \ref{lem:becCodes} below is our
main tool in quantifying the erasure decoding capabilities of the two
ensembles. Before stating the lemma, we mention a proposition showing
that linear condensers applied on \emph{affine sources} achieve either
zero or large errors:

\begin{prop} \label{prop:zeroError} Suppose that a distribution $\cX$
  is uniformly supported on an affine $m$-dimensional subspace over
  $\F_q^n$. Consider a linear function $f\colon \F_q^n \to \F_q^r$,
  and define the distribution $\cY$ as $\cY := f(\cX)$. Suppose that,
  for some $\eps < 1/2$, $\cY$ is $\eps$-close to
  having either min-entropy $r \log q$ or at least $m \log q$.  Then,
  $\eps=0$.
\end{prop}

\begin{proof}
  By linearity, $\cY$ is uniformly supported on an affine subspace $A$
  of $\F_q^r$. Let $m' \leq r$ be the dimension of this subspace, and observe
  that $m' \leq m$.
 
  First, suppose that $\cY$ is $\eps$-close to a distribution with
  min-entropy $r \log q$; i.e., the uniform distribution on $\F_q^r$.
  Now, the statistical distance between $\cY$ and the uniform
  distribution is, by definition, 
  \[
  \sum_{x \in A} ( q^{-m'} - q^{-r} ) = 1-q^{m'-r}.
  \]
  Since $\eps < 1/2$, $q \geq 2$, and $m'$ and $r$ are integers, this
  implies that the distance is greater than $1/2$ (a contradiction)
  unless $m'=r$, in which case it becomes zero.  Therefore, the output
  distribution is exactly uniform over $\F_q^r$.

  Now consider the case where $\cY$ is $\eps$-close to having
  min-entropy at least $m \log q$. Considering that $m' \leq m$, the
  definition of statistical distance implies that $\eps$ is at least
  \[
  \sum_{x \in A} (q^{-m'} - q^{-m}) = 1-q^{m'-m}.
  \]
  Similarly as before, we get that $m' = m$, meaning that $\cY$ is
  precisely a distribution with min-entropy $m \log q$.
\end{proof}

\begin{lem} \label{lem:becCodes} Let $S \subseteq [n]$ be a set of
  size at most $m$. Then all but a $3 \eps$ fraction of the codes in
  $\cF$ and all but a $3 \eps'$ fraction of those in $\cG$ can
  tolerate the erasure pattern defined by
  $S$. 
\end{lem}

\begin{proof}
  We prove the result for the ensemble $\cG$. The argument for $\cF$
  is similar.  Consider a bit-fixing source $\cS$ on $\F_2^n$
  that is uniform on the coordinates specified by $\bar{S} := [n]
  \setminus S$ and fixed to zeros elsewhere. Thus the min-entropy of
  $\cS$ is at least $n-m$, and the distribution $(U, g(\cS, U))$,
  where $U \sim \U_{d'}$, is $\eps'$-close to $\U_{d'+k}$.
  
  By Proposition~\ref{prop:strongExt}, for all but a $3 \eps'$ fraction
  of the choices of $u \in \F_2^{d'}$, the distribution of $g(\cS, u)$
  is $(1/3)$-close to $\U_k$. Fix such a $u$. By
  Proposition~\ref{prop:zeroError}, the distribution of $g(\cS, u)$
  must in fact be exactly uniform. Thus, the $k \times m$ submatrix of
  $G_u$ consisting of the columns picked by $\bar{S}$ must have rank
  $k$, which implies that for every $x \in \F_2^k$, the projection of
  the encoding $x \cdot G_u$ to the coordinates chosen by $\bar{S}$
  uniquely identifies $x$.
\end{proof}

The lemma combined with an simple averaging argument implies the following
corollary:

\begin{coro} \label{coro:becCodes} Let $\cS$ be any distribution on
  the subsets of $[n]$ of size at most $m$.  Then all but a
  $\sqrt{3\eps}$ (resp., $\sqrt{3\eps'}$) fraction of the codes in
  $\cF$ (resp., $\cG$) can tolerate erasure patterns sampled from
  $\cS$ with probability at least $1-\sqrt{3\eps}$ (resp.,
  $1-\sqrt{3\eps'}$), where the probability is taken over the
randomness of $\cS$ \qed
\end{coro}

Note that the result holds irrespective of the distribution $\cS$,
contrary to the familiar case of $\bec(p)$ for which the erasure
pattern is an i.i.d.\ (i.e., independent and identically-distributed)
sequence.  For the case of $\bec(p)$, the erasure pattern (regarded as
its binary characteristic vector in $\F_2^n$) is given by $S := (S_1,
\ldots, S_n)$, where the random variables $S_1, \ldots, S_n \in \F_2$
are i.i.d.\ and $\Pr[S_i = 1] = p$. We denote this particular
distribution by $\cB_{n,p}$, which assigns a nonzero probability to
every vector in $\F_2^n$.  Thus in this case we cannot directly apply
Corollary~\ref{coro:becCodes}. However, note that $\cB_{n,p}$ can be
written as a convex combination
\begin{equation} \label{eqn:convex} \cB_{n,p} = (1-\gamma) \U_{n,\leq
    p'} + \gamma \cD,
\end{equation}
for $p' := p + \Omega(1)$ that is arbitrarily close to $p$, where
$\cD$ is an ``error distribution'' whose contribution $\gamma$ is
exponentially small. The distribution $\U_{n, \leq p'}$ is the
distribution $\cB_{n,p}$ conditioned on vectors of weight at most
$np'$. Corollary~\ref{coro:becCodes} applies to $\U_{n,\leq p'}$ by
setting $m = np'$.  Moreover, by the convex combination above, the
erasure decoding error probability of any code for erasure pattern
distributions $\cB_{n,p}$ and $\U_{n,\leq p'}$ differ by no more than
$\gamma$. Therefore, the above result applied to the erasure
distribution $\U_{n,\leq p'}$ handles the particular case of $\bec(p)$
with essentially no change in the error probability.

In light of Corollary~\ref{coro:becCodes}, in order to obtain rates
arbitrarily close to the channel capacity, the output lengths of $f$
and $g$ must be sufficiently close to the entropy requirement
$m$. More precisely, it suffices to have $r \leq (1+\alpha) m$ and $k
\geq (1-\alpha) m$ for arbitrarily small constant $\alpha > 0$. The
seed length of $f$ and $g$ determine the size of the code ensemble.
Moreover, the error of the extractor and condenser determine the
erasure error probability of the resulting code ensemble. As achieving
the channel capacity is the most important concern for us, we will
need to instantiate $f$ (resp., $g$) with a linear lossless
condenser (resp., extractor) whose output length is close to $m$. We
mention suitable instantiations for each function.

For both functions $f$ and $g$, we can use the explicit extractor and
lossless condenser obtained from the Leftover Hash Lemma
(Lemma~\ref{lem:leftover}), which is optimal in the output length, but
requires a large seed, namely, $d=n$. The ensemble resulting this way
will thus have size $2^n$, but attains a positive error exponent
$\delta/2$ for an arbitrary rate loss $\delta > 0$.  Using an optimal
lossless condenser or extractor with seed length $d=\log(n) +
O(\log(1/\eps))$ and output length close to $m$, it is possible to
obtain a polynomially small capacity-achieving ensemble.  However, in
order to obtain an explicit ensemble of codes, the condenser of
extractor being used must be explicit as well.

In the world of linear extractors, we can use Trevisan's extractor
(Theorem~\ref{thm:RRV}) to improve the size of the ensemble compared
to what obtained from the Leftover Hash Lemma. In particular,
Trevisan's extractor combined with Corollary~\ref{coro:becCodes}
(using ensemble $\cG$) immediately gives the following result:

\begin{coro} \label{coro:bec} Let $p, c > 0$ be arbitrary
  constants. Then for every integer $n > 0$, there is an explicit
  ensemble $\cG$ of linear codes of rate $1-p-o(1)$ such that, the
  size of $\cG$ is quasipolynomial, i.e., $|\cG| = 2^{O(c^3 \log^3
    n)}$, and, all but an $n^{-c}=o(1)$ fraction of the codes in the
  ensemble have error probability at most $n^{-c}$ when used over
  $\bec(p)$. \qed
\end{coro}

For the ensemble $\cF$, on the other hand, we can use the linear
lossless condenser of Guruswami et al.\ that only requires a
logarithmic seed (Theorem~\ref{thm:GUV}).  Using this
condenser combined with Corollary~\ref{coro:becCodes}, we can
strengthen the above result as follows:

\begin{coro}
  \label{coro:becGUV} Let $p, c, \alpha > 0$ be arbitrary
  constants. Then for every integer $n > 0$, there is an explicit
  ensemble $\cF$ of linear codes of rate $1-p-\alpha$ such that $|\cG|
  = O(n^{c'})$ for a constant $c'$ only depending on $c, \alpha$.
  Moreover, all but an $n^{-c}=o(1)$ fraction of the codes in the
  ensemble have error probability at most $n^{-c}$ when used over
  $\bec(p)$. \qed
\end{coro}

\section{Codes for the Binary Symmetric Channel} \label{sec:bsc}

The goal of this section is to design capacity achieving code
ensembles for the binary symmetric channel $\bsc(p)$. In order to do
so, we obtain codes for the general (and not necessarily memoryless)
class $\SC(\F_q, \cZ)$ of symmetric channels, where $\cZ$ is any flat
distribution or sufficiently close to one.  For concreteness, we will
focus on the binary case where $q=2$. However, we remark that
the results and constructions can readily be extended to any fixed
prime power $q$.

Recall that the capacity of $\bsc(\cZ)$, seen as a binary channel, is
$1-h(\cZ)$ where $h(\cZ)$ is the entropy rate of $\cZ$.  The special
case $\bsc(p)$ is obtained by setting $\cZ = \cB_{n,p}$; i.e., the
product distribution of $n$ Bernoulli random variables with
probability $p$ of being equal to $1$.

The code ensemble that we use for the symmetric channel is the
ensemble $\cF$, obtained from linear lossless condensers, that we
introduced in the preceding section. Thus, we adopt the notation (and
parameters) that we used before for defining the ensemble $\cF$.
Recall that each code in the ensemble has rate at least $1-r/n$.  In
order to show that the ensemble is capacity achieving, we consider the
following brute-force decoder for each code:

\begin{quote}
  \emph{Brute-force decoder for code $\C_u$: } Given a received word
  $\hat{y} \in \F_2^n$, find a codeword $y \in \F_2^n$ of $\C_u$ used
  and a vector $z \in \supp(\cZ)$ such that $\hat{y} = y+z$. Output
  $y$, or an arbitrary codeword if no such pair is found. If there is
  more than one choice for the codeword $y$, arbitrarily choose one of
  them.
\end{quote}

For each $u \in \F_2^d$, denote by $\cE(\C_u, \cZ)$ the error
probability of the above decoder for code $\C_u$ over $\bsc(\F_2,
\cZ)$. The following lemma quantifies this probability:

\begin{lem} \label{lem:bscCodes} Let $\cZ$ be a flat distribution with
  entropy $m$.  Then for at least a $1-2\sqrt{\eps}$ fraction of the
  choices of $u \in \F_2^d$, we have $\cE(\C_u, \cZ) \leq
  \sqrt{\eps}$.
\end{lem}

\begin{proof}
  The intuitive idea behind the proof is that lossless condensers act
  ``almost injectively'' on the support of any input probability
  distribution with the prescribed entropy. We will use this property
  to construct a syndrome decoder for the code ensemble that achieves
  a sufficiently small error probability.
  
  By Proposition~\ref{prop:strongExt}, for a $1-2\sqrt{\eps}$ fraction
  of the choices of $u \in \zo^d$, the distribution $\cY := f(\cZ, u)$
  is $(\sqrt{\eps}/2)$-close to having min-entropy at least $m$. Fix
  any such $u$.  We show that the error probability $\cE(\C_u, \cZ)$
  is bounded by $\sqrt{\eps}$.
  
  For each $y \in \F_2^r$, define
  \[ \mathcal{N}(y) := | \{ x \in \supp(\cZ)\colon f(x, u) = y \}| \]
  and recall that $f(x, u) = H_u \cdot x$, where $H_u$ is a parity
  check matrix for $\C_u$. Now suppose that a message is encoded using
  the code $\C_u$ to an encoding $x \in \C_u$, and that $x$ is
  transmitted through the channel. The error probability $\cE(\C_u,
  \cZ)$ can be written as
  \begin{eqnarray}
    \cE(\C_u, \cZ) &=& \Pr_{z \sim \cZ}[ \exists x' \in \C_u, \exists z' \in \supp(\cZ)\setminus z \colon \nonumber \\ && \qquad \qquad x+z = x'+z' ] \nonumber \\
    &\leq& \Pr_{z \sim \cZ}[ \exists x' \in \C_u, \exists z' \in \supp(\cZ)\setminus z \colon \nonumber \\ && \qquad \qquad H_u\cdot (x+z) = H_u \cdot(x'+z') ]\nonumber \\ 
    &=& \Pr_{z \sim \cZ}[ \exists z' \in \supp(\cZ) \setminus z\colon \nonumber \\ && \qquad \qquad H_u\cdot z = H_u \cdot z' ] \label{eqn:errorProbCollision} \\
    &=& \Pr_{z \sim \cZ}[ \mathcal{N}(H \cdot z) > 1 ]\nonumber \\
    &=& \Pr_{z \sim \cZ}[ \mathcal{N}(f(x,u)) > 1 ], \label{eqn:errorProbLast}
  \end{eqnarray}
  where $\eqref{eqn:errorProbCollision}$ uses the fact that any
  codeword of $\C_u$ is in the right kernel of $H_u$.

  By the first part of Proposition~\ref{prop:flatmap}, there is a set
  $T \subseteq \F_2^r$ of size at least $(1-\sqrt{\eps})|\supp(\cZ)|$
  such that, $\mathcal{N}(y) = 1$ for every $y \in T$. Since $\cZ$ is
  uniformly distributed on its support, this combined with
  \eqref{eqn:errorProbLast} immediately implies that $\cE(\C_u, \cZ)
  \leq \sqrt{\eps}$.
\end{proof}

The lemma implies that any linear lossless condenser with entropy
requirement $m$ can be used to construct an ensemble of codes such
that all but a small fraction of the codes are good for reliable
transmission over $\bsc(\cZ)$, where $\cZ$ is an arbitrary flat
distribution with entropy at most $m$. Similar to the case of BEC, the
seed length determines the size of the ensemble, the error of the
condenser bounds the error probability of the decoder, and the output
length determines the proximity of the rate to the capacity of the
channel.
Again, using the condenser given by the Leftover Hash Lemma
(Lemma~\ref{lem:leftover}), we can obtain a capacity achieving
ensemble of size $2^n$. Moreover, using the linear lossless condenser
of Guruswami et al.\ (Theorem~\ref{thm:GUV}) the ensemble
can be made polynomially small (similar to the result given by
Corollary~\ref{coro:becGUV}).

It is not hard to see that the converse of the above result is also
true; namely, that any ensemble of linear codes that is universally
capacity achieving with respect to any choice of the noise
distribution $\cZ$ defines a linear lossless condenser. This
is spelled out in the lemma below.

\begin{lem} \label{lem:bscCodesConverse} Let $\{\C_1, \ldots, \C_T\}$
  be a binary code ensemble of length $n$ and dimension $n-r$ such
  that for every flat distribution $\cZ$ with min-entropy at most $m$
  on $\F_2^n$, all but a $\gamma$ fraction of the codes in the
  ensemble (for some $\gamma \in [0,1)$) achieve error probability at
  most $\eps$ (under syndrome decoding) when used over $\SC(\F_{q^n},
  \cZ)$. Then the function $f\colon \F_2^n \times [T] \to \F_2^{r}$
  defined as
  \[
  f(x,u) := H_u \cdot x,
  \]
  where $H_u$ is a parity check matrix for $\C_u$, is an
  $(m, 2\eps+\gamma)$-lossless condenser.
\end{lem}

\begin{proof}
  The proof is straightforward using similar arguments as in
  Lemma~\ref{lem:bscCodes}.  Without loss of generality (by a
  convexity argument), let $\cZ$ be a \emph{flat} distribution with
  min-entropy $m$, and denote by $D\colon \F_2^r \to \F_2^n$ the
  corresponding syndrome decoder. Moreover, without loss of generality
  we have taken the decoder to be a deterministic function. For a
  randomized decoder, one can fix the internal coin flips so as to
  preserve the upper bound on its error probability.  Now let $u$ be
  chosen such that $\C_u$ achieves an error probability at most $\eps$
  (we know this is the case for at least $\gamma T$ of the choices of
  $u$).
 
  Denote by $T \subseteq \supp(\cZ)$ the set of noise realizations
  that can potentially confuse the syndrome decoder. Namely,
  \[
  T := \{ z \in \supp(\cZ)\colon \exists z' \in \supp(\cZ), z' \neq z,
  H_u \cdot z = H_u \cdot z'\}.
  \]
  Note that, for a random $Z \sim \cZ$, conditioned on the event that
  $Z \in T$, the probability that the syndrome decoder errs on $Z$ is
  at least $1/2$, since we know that $Z$ can be confused by at least
  one different noise realization. We can write this more precisely as
  \[
  \Pr_{Z\sim \cZ}[ D(Z) \neq Z \mid Z \in T] \geq 1/2.
  \]
  Since the error probability of the decoder is upper bounded by
  $\eps$, we conclude that
  \[
  \Pr_{Z\sim \cZ}[ Z \in T] \leq 2\eps.
  \]
  Therefore, the fraction of the elements on support of $\cZ$ that
  collide with some other element under the mapping defined by $H_u$
  is at most $2\eps$. Namely,
  \[
  |\{ H_u \cdot z\colon z \in \supp(\cZ) \}| \geq 2^m(1-2\eps),
  \]
  and this is true for at least $1-\gamma$ fraction of the choices of
  $u$.  Thus, for a uniformly random $U \in [T]$ and $Z \sim \cZ$, the
  distribution of $(U, H_U \cdot Z)$ has a support of size at least
  \[
  (1-\gamma)(1-2\eps) T 2^m \geq (1-\gamma-2\eps) T 2^m.
  \]
  By the second part of Proposition~\ref{prop:flatmap}, we conclude
  that this distribution is $(2\eps+\gamma)$-close to having entropy
  $m+\log T$ and thus, the function $f$ defined in the statement is a
  lossless $(m, 2\eps+\gamma)$-condenser.
\end{proof}

By this lemma, any known lower bound on the seed length and the output
length of lossless condensers directly translates into lower bounds on
the size of the code ensemble and proximity to the capacity that can
be obtained from our framework.  In particular, it is known
\cite{ref:CRVW02} that any $(m,\eps)$-lossless condenser requires seed
length $d=\log(n/\eps)+\Omega(1)$ and has to output at least
$m+\log(1/\eps)+\Omega(1)$ bits.  These bounds are tight up to
additive constants, and are attained by random functions
\cite{ref:CRVW02}. Thus, in order to get codes with positive error
exponent in our framework (i.e., exponentially small error in the
block length), the size of the ensemble must be exponentially large.
Moreover, for any constant $c > 0$, the existence result of
\cite{ref:CRVW02} shows that there are lossless condensers that give
us ensembles of size $2^{cn}$ and positive error exponent (for all but
a $2^{\Omega(n)}$ fraction of the codes in the ensemble).

It is worthwhile to point out that the code ensembles $\cF$ and $\cG$
discussed in this and the preceding section preserve their erasure and
error correcting properties under any change of basis in the ambient
space $\F_2^n$, due to the fact that a change of basis applied on any
linear condenser results in a linear condenser with the same
parameters.  This is a property achieved by the trivial, but large,
ensemble of codes defined by the set of all $r \times n$ parity check
matrices.  Observe that no single code can be universal in this sense,
and it is inevitable to have a sufficiently large ensemble to attain
this property.

%
%

\subsection*{The Case $\bsc(p)$}

For the special case of $\bsc(p)$, the noise distribution $\cB_{n, p}$
is not a flat distribution. Fortunately, similar to the BEC case, we
can again use convex combinations to show that the result obtained in
Lemma~\ref{lem:bscCodes} can be extended to this important noise
distribution. The main tool that we need is an extension of
Lemma~\ref{lem:bscCodes} to convex combinations with a small number of
components.

Suppose that the noise distribution $\cZ$ is not a flat distribution
but can be written as a convex combination
\begin{equation} \label{eqn:noiseConvex} \cZ = \alpha_1 \cZ_1 + \cdots
  + \alpha_t \cZ_t.
\end{equation}
of $t$ flat distributions, where the number $t$ of summands is not too
large, and
\[
|\supp(\cZ_1)| \geq |\supp(\cZ_2)| \geq \cdots \geq |\supp(\cZ_t)|.
\]
For this more general case, we need to slightly tune our brute-force
decoder in the way it handles ties. In particular, we now require the
decoder to find a codeword $y \in \C_u$ and a potential noise vector
$z \in \supp(\cZ)$ that add up to the received word, as before.
However, in case more than one matching pair is found, we will require
the decoder to choose the one whose noise vector $z$ belongs to the
component $\cZ_1, \ldots, \cZ_t$ with smallest support (i.e., largest
index). If the noise vector $z \in \supp(\cZ_i)$ that maximizes the
index $i$ is still not unique, the decoder can arbitrarily choose one.
Under these conventions, we can now prove the following:

\begin{lem} \label{lem:bscConvex} Suppose that a noise distribution
  $\cZ$ is as in $\eqref{eqn:noiseConvex}$, where each component
  $\cZ_i$ has entropy at most $m$, and the function $f$ defining the
  ensemble $\cF$ is an $(m+1, \eps)$-lossless condenser.
  Then for at least a $1-t(t+1)\sqrt{\eps}$ fraction of the choices of
  $u \in \F_2^d$, the brute-force decoder satisfies $\cE(\C_u, \cZ)
  \leq 2t \sqrt{\eps}$.
\end{lem}

\begin{proof}
  For each $1 \leq i \leq j \leq t$, we define a flat distribution
  $\cZ_{ij}$ that is uniformly supported on $\supp(\cZ_i) \cup
  \supp(\cZ_j)$. Observe that each $\cZ_{ij}$ has min-entropy at most
  $m+1$ and thus the function $f$ is a lossless condenser with error
  at most $\eps$ for this source. By Proposition~\ref{prop:strongExt}
  combined with a union bound, for a $1-t(t+1)\sqrt{\eps}$ fraction of the
  choices of $u \in \zo^d$, all $t(t+1)/2$ distributions
  \[
  f(\cZ_{ij}, u)\colon 1 \leq i \leq j \leq t
  \]
  are simultaneously $(\sqrt{\eps}/2)$-close to having min-entropy at
  least $m$. Fix any such $u$.

  Consider a random variable $Z$, representing the channel noise, that
  is sampled from $\cZ$ as follows: First choose an index $I \in [t]$
  randomly according to the distribution induced by $(\alpha_1,
  \ldots, \alpha_t)$ over the indices, and then sample a random noise
  $Z \sim \cZ_I$.  Using the same line of reasoning leading to
  $\eqref{eqn:errorProbCollision}$ in the proof of
  Lemma~\ref{lem:bscCodes}, the error probability with respect to the
  code $\C_u$ (i.e., the probability that the tuned distance decoder
  gives a wrong estimate on the noise realization $Z$) can now be
  bounded as
  \begin{multline*}
  \cE(\C_u, \cZ) \leq \Pr_{I, Z}[ \exists i \in \{ I, \ldots, t\},\\
  \exists z' \in \supp(\cZ_i)\setminus Z\colon f(Z, u) = f(z', u) ].
  \end{multline*}
  For $i=1, \ldots, t$, denote by $\cE_i$ the right hand side
  probability in the above bound conditioned on the event that
  $I=i$. Fix any choice of the index $i$. Now it suffices to obtain an
  upper bound on $\cE_i$ irrespective of the choice of $i$, since
  \[
  \cE(\C_u, \cZ) \leq \sum_{i\in [t]} \alpha_i \cE_i.
  \]
  We call a noise realization $z \in \supp(\cZ_i)$ \emph{confusable}
  if
  \[
  \exists j \geq i, \exists z' \in \supp(\cZ_j) \setminus z\colon f(z,
  u) = f(z', u).
  \]
  That is, a noise realization is confusable if it can potentially
  cause the brute-force decoder to compute a wrong noise estimate. Our
  goal is to obtain an upper bound on the fraction of vectors on
  $\supp(\cZ_i)$ that are confusable.

  For each $j \geq i$, we know that $f(\cZ_{ij}, u)$ is
  $(\sqrt{\eps}/2)$-close to having min-entropy at least $m$.
  Therefore, by the first part of Proposition~\ref{prop:flatmap}, the
  set of confusable elements
  \begin{multline*}
  \{ z \in \supp(\cZ_i)\colon \exists z' \in \supp(\cZ_j)\setminus z \\
  \text{ such that } f(z, u)=f(z', u) \}
  \end{multline*}
  has size at most $\sqrt{\eps} |\supp(\cZ_{ij})| \leq 2\sqrt{\eps}
  |\supp(\cZ_i)|$ (using the fact that, since $j \geq i$, the support
  of $\cZ_j$ is no larger than that of $\cZ_i$).  By a union bound on
  the choices of $j$, we see that the fraction of confusable elements
  on $\supp(\cZ_i)$ is at most $2t\sqrt{\eps}$. Therefore, $\cE_i \leq
  2t\sqrt{\eps}$ and we get the desired upper bound on the error
  probability of the brute-force decoder.
\end{proof}

The result obtained by Lemma~\ref{lem:bscConvex} can be applied to the
channel $\bsc(p)$ by observing that the noise distribution $\cB_{n,p}$
can be written as a convex combination
\[
\cB_{n,p} = \sum_{i=n(p-\eta)}^{n(p+\eta)} \alpha_i \U_{n,i} + \gamma
\cD,
\]
where $\U_{n,i}$ denotes the flat distribution supported on binary
vectors of length $n$ and Hamming weight exactly $i$, and $\cD$ is the
distribution $\cB_{n,p}$ conditioned on the vectors whose Hamming
weights lie outside the range $[n(p-\eta), n(p+\eta)]$.  The parameter
$\eta > 0$ can be chosen as an arbitrarily small real number, so that
the min-entropies of the distributions $\U_{n,i}$ become arbitrarily
close to the Shannon entropy of $\cB_{n,p}$; namely, $n h(p)$. This
can be seen by the estimate
\[
\binom{n}{w} = 2^{n h(w/n) \pm o(n)},
\]
$h(\cdot)$ being the binary entropy function, that is easily derived
from Stirling's formula.  By Chernoff bounds, the error $\gamma$ can
be upper bounded as
\[
\gamma = \Pr_{Z \sim \cB_{n,p}}[|\wgt(Z) - np| > \eta n] \leq 2
e^{-c_\eta np} = 2^{-\Omega(n)},
\]
where $c_\eta > 0$ is a constant only depending on $\eta$, and is thus
exponentially small.  Thus the error probability attained by any code
under noise distributions $\cB_{n,p}$ and $\cZ :=
\sum_{i=n(p-\eta)}^{n(p+\eta)} \alpha_i \U_{n,i}$ differ by the
exponentially small quantity $\gamma$. We may now apply
Lemma~\ref{lem:bscConvex} on the noise distribution $\cZ$ to attain
code ensembles for the binary symmetric channel $\bsc(p)$.  The error
probability of the ensemble is at most $2n\sqrt{\eps}$, and this bound
is satisfied by at least a $1-n^2 \sqrt{\eps}$ fraction of the codes.

Finally, the code ensemble is capacity achieving for $\bsc(p)$
provided that the condenser $f$ attains an output length $r \leq
(1+\alpha)(p+\eta)n$ for arbitrarily small constant $\alpha$, and
$\eps = o(n^{-4})$. Same as before, the required bounds on the output
length and error are in particular attained by the Leftover Hash Lemma
(Lemma~\ref{lem:leftover}) and the lossless condenser of Guruswami et
al.~(Theorem~\ref{thm:GUV}).  The parameters achieved by the resulting
explicit ensembles are summarized in the table below. These are
essentially the same as what we could get for the BEC and $\bsc(\cZ)$
channels before.

\begin{center}
\noindent \begin{tabular}{|l|l|l|} \hline
Condenser used & Ensemble size & Error probability \\
\hline
Lemma \ref{lem:leftover} & $2^n$ & $2^{-\Omega(n)}$ \\
Theorem \ref{thm:GUV} & $\poly(n)$ & $n^{-c}\ (\forall c > 0)$ \\
\hline 
\end{tabular}
\end{center}

\section{Explicit Capacity Achieving Codes} \label{sec:explicit}

In the preceding sections, we showed how to obtain small ensembles of
explicit capacity achieving codes for various discrete channels,
including the important special cases $\bec(p)$ and $\bsc(p)$.  Two
drawbacks related to these constructions are:
\begin{enumerate}
\item While an overwhelming fraction of the codes in the ensemble are
  capacity achieving, in general it is not clear how to pin down a
  single, capacity achieving code in the ensemble.
 
\item For the symmetric additive noise channels, the brute-force
  decoder is extremely inefficient and is of interest only for proving
  that the constructed ensembles are capacity achieving.
\end{enumerate}

In a classic work, Justesen \cite{ref:Justesen} showed that the idea
of code concatenation
first introduced by Forney \cite{ref:forney} can be used to transform
any ensemble of capacity achieving codes, for a memoryless channel,
into an explicit, efficiently decodable code with improved error
probability over the same channel. In this section we revisit this
idea and apply it to our ensembles. For concreteness, we focus on the
binary case and consider a memoryless channel $\mathscr{C}$ that is
either $\bec(p)$ or $\bsc(p)$.

Throughout this section, we consider an ensemble $\cS$ of linear codes
with block length $n$ and rate $R$, for which it is guaranteed that
all but a $\gamma = o(1)$ fraction of the codes are capacity achieving
(for a particular discrete memoryless symmetric channel, in our case
either $\bec(p)$ or $\bsc(p)$) with some vanishing error probability
$\eta = o(1)$ (the asymptotics are considered with respect to the
block length $n$).

\newcommand{\cout}{\C_{\mathrm{out}}} Justesen's concatenated codes
take an outer code $\cout$ of block length $s := |\cS|$, alphabet
$\F_{2^k}$, rate $R'$ as the \emph{outer code}. The particular choice
of the outer code in the original construction is Reed-Solomon codes.
However, we point out that any outer code that allows unique decoding
of some constant fraction of errors at rates arbitrarily close to one
would suffice for the purpose of constructing capacity achieving
codes. In particular, in this section we will use an expander-based
construction of asymptotically good codes due to Spielman
\cite{ref:Spielman}, from which the following theorem can be easily
derived\footnote{There are alternative choices of the outer code that
  lead to a similar result, e.g., expander-based codes due to
  Guruswami and Indyk \cite{ref:GI05}.}:

\begin{thm} \label{thm:expander} For every integer $k > 0$ and every
  absolute constant $R' < 1$, there is an explicit family of
  $\F_2$-linear codes over $\F_{2^k}$ for every block length and rate
  $R'$ that is error-correcting for an $\Omega(1)$ fraction of
  errors. The running time of the encoder and the decoder is linear in
  the bit-length of the codewords.
\end{thm}

\subsection{Justesen's Concatenation Scheme} \index{Justesen's
  concatenation}

The concatenation scheme of Justesen differs from traditional
concatenation in that the outer code is concatenated with an ensemble
of codes rather than a single inner code.

In this construction, size of the ensemble is taken to be matching
with the block length of the outer code, and each symbol of the outer
code is encoded with one of the inner codes in the ensemble. We use
the notation $\C := \cout \diamond \cS$ to denote concatenation of an
outer code $\cout$ with the ensemble $\cS$ of inner codes.  Suppose
that the alphabet size of the outer code is taken to be $2^{\lfloor Rn
  \rfloor}$, where we recall that $n$ and $R$ denote the block length
and rate of the inner codes in $\cS$.

\begin{figure*}
  \centerline{\includegraphics[width=\textwidth-5cm]{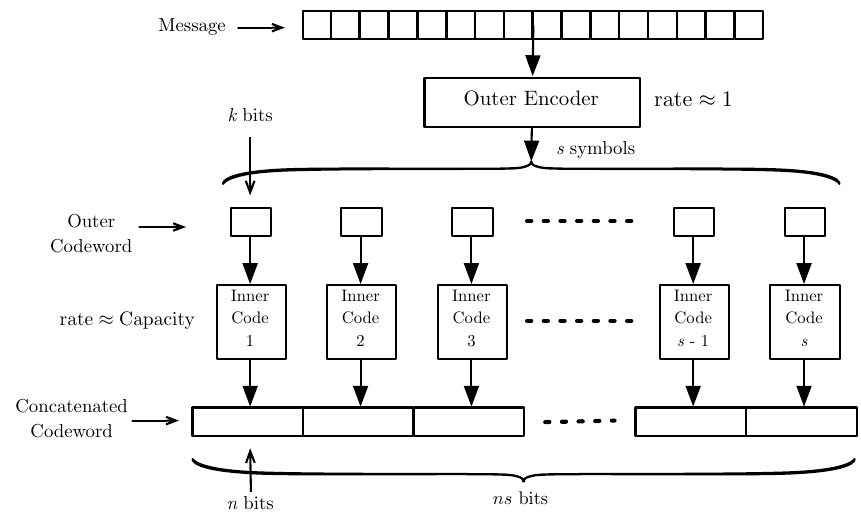}}
  \caption[Justesen's concatenation scheme]{Justesen's concatenation
    scheme.}
  \label{fig:justesen}
\end{figure*}

The encoding of a message with the concatenated code can be obtained
as follows: First, the message is encoded using $\cout$ to obtain an
encoding $(c_1, \ldots, c_{s}) \in \F_{2^k}^s$, where $k = \lfloor Rn
\rfloor$ denotes the dimension of the inner codes. Then, for each $i
\in [s]$, the $i$th symbol of the encoding $c_i$ is further encoded by
the $i$th code in the ensemble $\cS$ (under some arbitrary ordering of
the codes in the ensemble), resulting in a binary sequence $c'_i$ of
length $n$. The $ns$-bit long binary sequence $(c'_1, \ldots, c'_s)$
defines the encoding of the message under $\cout \diamond \cS$.  The
concatenation is scheme is depicted in Fig.~\ref{fig:justesen}.

Similar to classical concatenated codes, the resulting binary code
$\C$ has block length $N := n s$ and dimension $K := kk'$, where $k'$
is the dimension of the outer code $\cout$.  However, the neat idea in
Justesen's concatenation is that it eliminates the need for a
brute-force search for finding a good inner code, as long as almost
all inner codes are guaranteed to be good.

\subsection{The Analysis}

In order to analyze the error probability attained by the concatenated
code $\cout \diamond \cS$, we consider the following naive
decoder\footnote{Alternatively, one could use methods such as Forney's
  Generalized Minimum Distance (GMD) decoder for Reed-Solomon codes
  \cite{ref:forney}.  However, the naive decoder suffices for our
  purposes and works for any asymptotically good choice of the outer
  code.}:

\begin{enumerate}
\item Given a received sequence $(y_1, \ldots, y_s) \in (\F_2^n)^{s}$,
  apply an appropriate decoder for the inner codes (e.g., the
  brute-force decoder for BSC, or Gaussian elimination for BEC) to
  decode each $y_i$ to a codeword $c'_i$ of the $i$th code in the
  ensemble.
 
\item Apply the outer code decoder on $(c'_1, \ldots, c'_{s})$ that is
  guaranteed to correct some constant fraction of errors, to obtain a
  codeword $(c_1, \ldots, c_{s})$ of the outer code $\cout$.
 
\item Recover the decoded sequence from the corrected encoding $(c_1,
  \ldots, c_{s})$.
\end{enumerate}

Since the channel is assumed to be memoryless, the noise distributions
on inner codes are independent. Let $\cG \subseteq [s]$ denote the set
of coordinate positions corresponding to ``good'' inner codes in $\cS$
that achieve an error probability bounded by $\eta$.  By assumption,
we have $\cG \geq (1-\gamma)|\cS|$.

Suppose that the outer code $\cout$ corrects some $\gamma + \alpha$
fraction of adversarial errors, for a constant $\alpha > \eta$.  Then
an error might occur only if more than $\alpha N$ of the codes in
$\cG$ fail to obtain a correct decoding. We expect the number of
failures within the good inner codes to be $\eta |\cG|$. Due to the
noise independence, it is possible to show that the fraction of
failures may deviate from the expectation $\eta$ only with a
negligible probability. In particular, a direct application of
the Chernoff bounds implies that the probability that more than an $\alpha$
fraction of the good inner codes err is at most
\begin{equation} \label{eqn:justesenError} \eta^{\alpha' |\cG|} =
  2^{-\Omega_\alpha(\log(1/\eta) s)},
\end{equation}
where $\alpha' > 0$ is a constant that only depends on $\alpha$.  This
also upper bounds the error probability of the concatenated code.  In
particular, we see that if the error probability $\eta$ of the inner
codes is exponentially small in their block length $n$, the
concatenated code also achieves an exponentially small error in its
block length $N$.

Now we analyze the encoding and decoding complexity of the
concatenated code, assuming that Spielman's expander codes
(Theorem~\ref{thm:expander}) are used for the outer code. With this
choice, the outer code becomes equipped with a linear-time encoder and
decoder.  Since any linear code can be encoded in quadratic time (in
its block length), the concatenated code can be encoded in $O(n^2 s)$,
which for $s \gg n$ can be considered ``almost linear'' in the block
length $N = ns$ of $\C$. The decoding time of each inner code is cubic
in $n$ for the erasure channel, since decoding reduces to Gaussian
elimination, and thus for this case the naive decoder runs in time
$O(n^3 s)$.  

For the symmetric channel, however, the brute-force
decoder used for the inner codes takes exponential time in the block
length, namely, $2^{Rn} \poly(n)$.  Therefore, the running time of the
decoder for concatenated code becomes bounded by $O(2^{Rn} s
\poly(n))$. When the inner ensemble is exponentially large; i.e., $s =
2^n$ (which is the case for our ensembles if we use the Leftover Hash
Lemma), the decoding complexity becomes $O(s^{1+R} \poly(\log s))$
which is at most quadratic in the block length of $\C$.

Since the rate $R'$ of the outer code can be made arbitrarily close to
$1$ (while keeping the minimum distance linear), rate of the
concatenated code $\C$ can be made arbitrarily close to the rate $R$
of the inner codes. Thus, if the ensemble of inner codes is
capacity-achieving, so would be the concatenated code.

\subsection{Density of the Explicit Family} \label{sec:density}

In the preceding section we saw how to obtain explicit capacity
achieving codes from capacity achieving code ensembles using
concatenation.  One of the important properties of the resulting
family of codes that is influenced by the size of the inner code
ensemble is the set of block lengths $N$ for which the concatenated
code is defined.  Recall that $N = ns$, where $n$ and $s$ respectively
denote the block length of the inner codes and the size of the code
ensemble, and the parameter $s$ is a function of $n$. For instance,
for all classical examples of capacity achieving code ensembles
(namely, Wozencraft's ensemble, Goppa codes and shortened cyclic
codes) we have $s(n) = 2^n$. In this case, the resulting explicit
family of codes would be defined for integer lengths of the form $N(i)
= i 2^i$.

A trivial approach for obtaining capacity achieving codes for all
lengths is to use a \emph{padding trick}. Suppose that we wish to
transmit a particular bit sequence of length $K$ through the channel
using the concatenated code family of rate $\rho$ that is taken to be
sufficiently close to the channel capacity. The sequence might
originate from a source that does not produce a constant stream of
bits (e.g., consider a terminal emulator that produces data only when
user input is available).

Ideally, one requires the length of the encoded sequence to be $N =
\lceil K/\rho \rceil$. However, since the family might not be defined
for the block length $N$, we might be forced to take a code $\C$ in
the family with smallest length $N' \geq N$ that is of the form $N' =
n s(n)$, for some integer $n$, and pad the original message with
redundant symbols.  This way we have encoded a sequence of length $K$
to one of length $N'$, implying an effective rate $K/N'$.  The rate
loss incurred by padding is thus equal to $\rho - K/N' = K(1/N -
1/N')$.  Thus, if $N' \geq N(1+\delta)$ for some positive constant
$\delta > 0$, the rate loss becomes lower bounded by a constant and
subsequently, even if the original concatenated family is capacity achieving,
it no longer remains capacity achieving when extended to arbitrarily
chosen lengths using the padding trick.

Therefore, if we require the explicit family obtained from
concatenation to remain capacity achieving for all lengths, the set of
block lengths $\{i \cdot s(i)\}_{i \in \N}$ for which it is defined must be
sufficiently dense. This is the case provided that we have
\[
\frac{s(n)}{s(n+1)} = 1-o(1),
\]
which in turn, requires the capacity achieving code ensemble to have a
sub-exponential size (by which we mean $s(n) = 2^{o(n)}$).

Using the framework introduced in this paper, linear extractors and
lossless condensers that achieve nearly optimal parameters would
result in code ensembles of polynomial size in $n$.  The explicit
erasure code ensemble obtained from Trevisan's extractor
(Corollary~\ref{coro:bec}) or Guruswami-Umans-Vadhan's lossless
condenser (Corollary~\ref{coro:becGUV}) combined with Justesen's
concatenation scheme results in an explicit sequence of capacity
achieving codes for the binary erasure channel that is defined for
every block length, and allows almost linear-time (i.e., $N^{1+o(1)}$)
encoding and decoding. Moreover, the latter sequence of codes that is
obtained from a lossless condenser is capacity achieving for the
binary symmetric channel (with a matching bit-flip probability) as
well.

\section{Duality of Linear Affine
  Condensers} \label{sec:dualityAffine}

In Section~\ref{sec:bec} we saw that linear extractors for bit-fixing
sources can be used to define generator matrices of a family of
erasure-decodable codes. On the other hand, we showed that linear
lossless condensers for bit-fixing sources define parity check
matrices of erasure-decodable codes.

Recall that generator and parity check matrices are dual notions, and
in our construction we have considered matrices in one-to-one
correspondence with linear mappings.  Indeed, we have used linear
mappings defined by extractors and lossless condensers to obtain
generator and parity check matrices of our codes (where the $i$th row
of the matrix defines the coefficient vector of the linear form
corresponding to the $i$th output of the mapping).  Thus, we get a
natural duality between linear functions: If two linear functions
represent generator and parity check matrices of the same code, they
can be considered dual\footnote{Note that, under this notion of
  duality, the dual of a linear function need not be unique even
  though its linear-algebraic properties (e.g., kernel) would be
  independent of its choice.}.  Just in the same way that the number
of rows of a generator matrix and the corresponding parity check
matrix add up to their number of columns (provided that there is no
linear dependence between the rows), the dual of a linear function
mapping $\F_q^n$ to $\F_q^k$ (where $k \leq n$) that has no linear
dependencies among its $n-k$ outputs can be taken to be a linear
function mapping $\F_q^n$ to $\F_q^{n-k}$.

In fact, a duality between linear extractors and lossless condensers
for affine sources is implicit in the analysis leading to
Corollary~\ref{coro:becCodes}. Namely, it turns out that if a linear
function is an extractor for an affine source, the dual function
becomes a \emph{lossless condenser} for the \emph{dual distribution},
and vice versa.  This is made precise (and slightly more general) in
the following theorem.

\begin{thm} \label{thm:conDuality} Suppose that the linear mapping
  defined by a matrix $G \in \F_q^{r\times n}$ of rank $r \leq n$ is
  an $(m \log q) \to_\eps (m' \log q)$ condenser for an
  $m$-dimensional affine source $\cX$ over $\F_q^n$ and $\eps < 1/2$
  so that for $X \sim \cX$, the distribution of $G \cdot X^\top$ is
  $\eps$-close to having min-entropy at least $m' \log q$.  Let $H \in
  \F_q^{(n-r) \times n}$ be a dual matrix for $G$ (i.e., $G H^\top =
  0$) of rank $n-r$ and $\cY$ be an $(n-m)$-dimensional affine space
  over $\F_q^n$ supported on any translation of the dual subspace
  corresponding to the support of $\cX$. Then for $Y \sim \cY$, the
  distribution of $H \cdot Y^\top$ has entropy at least $(n-m+m'-r)
  \log q$.
\end{thm}

\begin{proof}
  In light of Proposition~\ref{prop:zeroError}, without loss of
  generality we may assume that $\eps = 0$, and thus, the distribution
  of $G \cdot X^\top$ has min-entropy at least $m' \log q$.

  \newcommand{\rker}{\mathsf{rker}} \newcommand{\lker}{\mathsf{lker}}
Suppose that $\cX$ is supported on a set
  \[
  \{ x \cdot A_G + a\colon x \in \F_q^m \},
  \]
  where $A_G \in \F_q^{m \times n}$ has rank $m$ and $a \in \F_q^n$ is
  a fixed row vector.  Moreover we denote the dual distribution $\cY$
  by the set
  \[
  \{ y \cdot A_H + b\colon y \in \F_q^{n-m} \},
  \]
  where $b \in \F_q^n$ is fixed and $A_H \in \F_q^{(n-m) \times n}$ is
  of rank $n-m$, and we have the orthogonality relationship $A_H \cdot
  A_G^\top = 0$.

  The assumption that $G$ is an $(m \log q) \to_0 (m' \log
  q)$-condenser implies that the distribution
  \[
  G \cdot (A_G^\top \cdot \U_{\F_q^m} + a^\top),
  \]
  where $\U_{\F_q^m}$ stands for a uniformly random row vector in
  $\F_q^m$, is an affine source of dimension at least $m'$, equivalent
  to saying that the matrix $G \cdot A_G^\top \in \F_q^{r \times m}$
  has rank at least $m'$ (since rank is equal to the dimension of the
  image), or in symbols,
  \begin{equation} \label{eqn:rankGAg} \rk(G \cdot A_G^\top) \geq m'.
  \end{equation}
  Observe that since we have assumed $\rk(G) = r$, its right kernel is
  $(n-r)$-dimensional, and thus the linear mapping defined by $G$
  cannot reduce more than $n-r$ dimensions of the affine source
  $\cX$. Thus, the quantity $n-m+m'-r$ is non-negative.

  By a similar argument as above, in order to show the claim we need
  to show that
  \[
  \rk(H \cdot A_H^\top) \geq n-m+m'-r.
  \]
  Suppose not. Then the right kernel of $H\cdot A_H^\top \in
  \F_q^{(n-r)\times(n-m)}$ must have dimension larger than
  $(n-m)-(n-m+m'-r) = r-m'$. Denote this right kernel by $\mathcal{R}
  \subseteq \F_q^{n-m}$.  Since the matrix $A_H$ is assumed to have
  maximal rank $n-m$, and $n-m \geq r-m'$, for each nonzero $y \in
  \mathcal{R}$, the vector $y \cdot A_H \in \F_q^n$ is nonzero and
  since $H \cdot (A_H^\top y^\top) = 0$ (by the definition of right
  kernel), the duality of $G$ and $H$ implies that there is a nonzero
  $x \in \F_q^r$ where
  \[
  x \cdot G = y \cdot A_H,
  \]
  and the choice of $y$ uniquely specifies $x$. In other words, there
  is a subspace $\mathcal{R'} \subseteq \F_q^r$ such that \[
  \dim(\mathcal{R'}) = \dim(\mathcal{R}), \] and
  \[
  \{ x \cdot G\colon x \in \mathcal{R'} \} = \{ y \cdot A_H\colon y
  \in \mathcal{R} \}.
  \]
  But observe that, by orthogonality of $A_G$ and $A_H$, every $y$
  satisfies $y \cdot A_H A_G^\top = 0$, meaning that for every $x \in
  \mathcal{R'}$, we must have $x \cdot G A_G^\top = 0$.  Thus,
\[
\dim(\mathsf{LeftKernel}(G A_G^\top)) \geq \dim(\mathcal{R}') = \dim(\mathcal{R}) > r-m',
\]
which implies, for the $r \times m$ matrix $G A_G^\top$, that
\[
\rk(G A_G^\top) = r - \dim(\mathsf{LeftKernel}(G A_G^\top)) < m',
\]
which is a contradiction for \eqref{eqn:rankGAg}.
\end{proof}

Two important special cases of the above result are related to affine
extractors ($m'=r$) and lossless condensers ($m' = m$). When the
linear mapping $G$ is an affine extractor for an $m$-dimensional
affine source $A$, the dual mapping $H$ becomes a lossless condenser
for the $(n-m)$-dimensional affine source supported on any translation
of the dual subspace $A^\top$ corresponding to $A$, and vice versa.

Moreover, we immediately get a duality theorem for \emph{seeded}
affine condensers as well. A seeded $m \to_\eps m'$ affine condenser
is a function $f\colon \F_q^n \times \zo^d \to \F_q^r$ that is
guaranteed to satisfy the requirements of Definition~\ref{def:condenser}
only for affine sources. Linear seeded affine condensers are
particularly interesting objects in derandomization theory, especially
as building blocks for construction of seedless affine extractors
\cite{ref:affine}. For a seeded condenser, the dual function is,
naturally, any seeded function $g\colon \F_q^n \times \zo^d \to
\F_q^{n-r}$ such that for every seed $z \in \zo^d$, the functions
$g(\cdot, z)$ and $f(\cdot, z)$ are dual linear functions.

Using the notions above and Theorem~\ref{thm:conDuality},
we conclude the following:

\begin{coro} \label{coro:conDuality} Let $f\colon \F_q^n \times \zo^d
  \to \F_q^r$ and $g\colon \F_q^n \times \zo^d \to \F_q^{n-r}$ be dual
  seeded functions\footnote{ We have implicitly assumed, without loss
    of generality, that for every fixed seed $z \in \zo^d$, the linear
    functions $f(\cdot, z)$ and $g(\cdot, z)$ are surjective.}.  Then,
  for every $\eps < 1/2$, and integers $m' \leq m \leq n$ (where $m'
  \leq r$), the function $f$ is an $(m \log q) \to_\eps (m' \log q)$
  condenser for affine sources if any only if $g$ is an $(n-m) \log q
  \to_\eps (n-m+m'-r) \log q$ condenser for affine sources. In
  particular, $f$ is an $(m, \eps)$ affine extractor if and only if
  $g$ is an $(n-m, \eps)$-lossless condenser for affine sources.  \qed
\end{coro}

\bibliographystyle{IEEEtran} \bibliography{full}

\appendix

\subsection{Proof of Proposition~\ref{prop:flatmap}}
\label{app:flatmap}

  Suppose that $\cX$ is uniformly supported on a set $S \subseteq
  \Omega$ of size $M$, and denote by $\mu$ the distribution $f(\cX)$
  over $\Gamma$. For each $y \in \Gamma$, define \[n_y := |\{x \in
  \supp(\cX)\colon f(x) = y\}|.\] Moreover, define $T := \{ y \in
  \Gamma\colon n_y = 1\}$, and similarly, $T' := \{ y \in \Gamma\colon
  n_y \geq 2\}$. Observe that for each $y \in \Gamma$ we have $\mu(y)
  = n_i/M$, and also $\supp(\mu) = T \cup T'$.  Thus,
  \begin{equation} \label{eqn:TTp} |T| + \sum_{y \in T'} n_y = M.
  \end{equation}

  Now we show the first assertion. Denote by $\mu'$ a distribution on
  $\Gamma$ with min-entropy $M$ that is $\eps$-close to $\mu$, which
  is guaranteed to exist by the assumption.  The fact that $\mu$ and
  $\mu'$ are $\eps$-close implies that
  \[
  \sum_{y \in T'} | \mu(y) - \mu'(y) | \leq \eps \Rightarrow \sum_{y
    \in T'} (n_y - 1) \leq \eps M.
  \]
  In particular, this means that $|T'| \leq \eps M$ (since by the
  choice of $T'$, for each $y \in T'$ we have $n_y \geq 2$).
  Furthermore,
  \[
  \sum_{y \in T'} (n_y - 1) \leq \eps M \Rightarrow \sum_{y \in T'}
  n_y \leq \eps M + |T'| \leq 2\eps M.
  \]
  This combined with \eqref{eqn:TTp} gives
  \[
  |T| = M - \sum_{y \in T'} n_y \geq (1-2\eps) M
  \]
  as desired.

  For the second part, observe that $|T'| \leq \eps M$. Let $\mu'$ be
  any flat distribution with a support of size $M$ that contains the
  support of $\mu$. The statistical distance between $\mu$ and $\mu'$
  is equal to the difference between the probability mass of the two
  distributions on those elements of $\Gamma$ to which $\mu'$ assigns
  a bigger probability, namely,
  \begin{eqnarray*}
  \frac{1}{M}(\supp(\mu') - \supp(\mu)) &=& \frac{\sum_{y \in T'}
    (n_y-1)}{M} \\ &=& \frac{\sum_{y \in T'} n_y - |T'|}{M} \\ &=&
  \frac{M-|T|-|T'|}{M},
  \end{eqnarray*}
  where we have used \eqref{eqn:TTp} for the last equality.  But
  $|T|+|T'| = |\supp(\mu)| \geq (1-\eps)M$, giving the required bound. \qed

  \subsection{Proof of Lemma~\ref{lem:leftover}}
  \label{app:LOHL}

  This proof is based on a proof of the original Leftover Hash Lemma
  in \cite{ref:IZ89}.  It is easy to see and well known that any
  distribution with min-entropy at least $m$ is a convex combination
  of \emph{flat} distributions with min-entropy $m$; that is,
  distributions that are uniformly supported on a set of size $M :=
  2^m$. Thus, it is sufficient to prove the lemma for a flat
  distribution $\cX$ supported on a set $\supp(\cX)$ of size $M$.

  Define $R := 2^r$, $D := 2^d$, and let $\mu$ be any flat
  distribution over $\F_2^{d+r}$ such that $\supp(\cX) \subseteq
  \supp(\mu)$, and denote by $\cY$ the distribution of $(Z, f(X, Z))$
  over $\F_2^{d+r}$ where $X \sim \cX$ and $Z \sim \U_d$.  We will
  first upper bound the $\ell_2$ distance of the two distributions
  $\cY$ and $\mu$, that can be expressed as follows:
  {\allowdisplaybreaks
    \begin{eqnarray}
      \| \cY - \mu \|_2^2 &=& \sum_{x \in \F_2^{d+r}} (\cY(x) - \mu(x))^2 \nonumber \\
      &=& \sum_{x} \cY(x)^2 + \sum_{x} \mu(x)^2 -2 \sum_{x} \cY(x) \mu(x) \nonumber \\
      &\stackrel{\mathrm{(a)}}{=}& \sum_{x} \cY(x)^2 + \nonumber \\ && \qquad \frac{1}{|\supp(\mu)|} -\frac{2}{|\supp(\mu)|} \sum_{x} \cY(x) \nonumber \\
      &=& \sum_{x} \cY(x)^2 - \frac{1}{|\supp(\mu)|} \label{eqn:LLLa},
    \end{eqnarray}}
  where $\mathrm{(a)}$ uses the fact that $\mu$ assigns probability $1/|\supp(\mu)|$
  to exactly $|\supp(\mu)|$ elements of $\F_2^{d+r}$ and zeros elsewhere.

  Now observe that $\cY(x)^2$ is the probability that two independent
  samples drawn from $\cY$ turn out to be equal to $x$, and thus,
  $\sum_{x} \cY(x)^2$ is the \emph{collision probability} of two
  independent samples from $\cY$, which can be written as
  \begin{equation*}
    \sum_{x} \cY(x)^2 = \Pr_{Z,Z',X,X'}[(Z, f(X, Z)) = (Z', f(X', Z'))],
  \end{equation*}
  where $Z, Z' \sim \F_2^d$ and $X, X' \sim \cX$ are independent
  random variables.  We can rewrite the collision probability as
  {\allowdisplaybreaks
    \begin{eqnarray*}
      \sum_{x} \cY(x)^2 &=& \Pr[Z = Z'] \times \\ && \qquad  \Pr[ f(X, Z) = f(X', Z') \mid Z = Z'] \\
      &=& \frac{1}{D} \cdot \Pr_{Z,X,X'}[ h_{Z}(X) = h_{Z}(X') ] \\
      &=& \frac{1}{D} \cdot (\Pr[X=X'] + \\ && \frac{1}{M^2} \sum_{\substack{x, x' \in \supp(\cX) \\ x \neq x'}} \Pr_{Z}[ h_{Z}(x) = h_{Z}(x') ]) \\
      &\stackrel{\mathrm{(b)}}{\leq}& \frac{1}{D} \cdot \big(\frac{1}{M} + \frac{1}{M^2} \sum_{\substack{x, x' \in \supp(\cX) \\ x \neq x'}} \frac{1}{R}\big) \\
      &\leq& \frac{1}{DR} \cdot \big(1 + \frac{R}{M}\big),
    \end{eqnarray*}}
  \noindent where $\mathrm{(b)}$ uses the assumption that $\cH$ is a
  universal hash family.  Plugging the bound in \eqref{eqn:LLLa}
  implies that
  \[
  \| \cY - \mu \|_2 \leq \frac{1}{\sqrt{DR}} \cdot \sqrt{1 -
    \frac{DR}{|\supp(\mu)|} + \frac{R}{M}}.
  \]
  Observe that both $\cY$ and $\mu$ assign zero probabilities to
  elements of $\F_2^{d+r}$ outside the support of $\mu$. Thus using
  the Cauchy-Schwarz inequality on a domain of size $\supp(\mu)$, the
  above bound implies that the statistical distance between $\cY$ and
  $\mu$ is at most
  \begin{equation} \label{eqn:LLLb} \frac{1}{2} \cdot
    \sqrt{\frac{|\supp(\mu)|}{DR}} \cdot \sqrt{1 -
      \frac{DR}{|\supp(\mu)|} + \frac{R}{M}}.
  \end{equation}
  Now, for the first part of the lemma, we specialize $\mu$ to the
  uniform distribution on $\F_2^{d+r}$, which has a support of size
  $DR$, and note that by the assumption that $r \leq m -
  2\log(1/\eps)$ we have $R \leq \eps^2 M$.  Using \eqref{eqn:LLLb},
  it follows that $\cY$ and $\mu$ are $(\eps/2)$-close.

  On the other hand, for the second part of the theorem, we specialize
  $\mu$ to any flat distribution on a support of size $DM$ containing
  $\supp(\cY)$ (note that, since $\cX$ is assumed to be a flat
  distribution, $\cY$ must have a support of size at most $DM$). Since
  $r \geq m + 2\log(1/\eps)$, we have $M = \eps^2 R$, and again
  \eqref{eqn:LLLb} implies that $\cY$ and $\mu$ are $(\eps/2)$-close.
  \qed

\end{document}